\documentclass[journal]{IEEEtran}
\usepackage{amsmath,amsthm}
\usepackage{epsfig}
\usepackage{graphicx}
\usepackage[hyphens]{url}
\usepackage{color}
\usepackage[colorlinks=true,citecolor=black,urlcolor=black,linkcolor=black,hyperindex,breaklinks]{hyperref}
\usepackage[table,hyperref,x11names]{xcolor}
\usepackage{comment}
\usepackage{array}
\usepackage{hhline}
\usepackage[ruled,vlined,linesnumbered]{algorithm2e}
\usepackage{pifont}
\usepackage{amssymb}
\usepackage{mathrsfs}
\usepackage{mathcomp}
\usepackage{multirow}
\usepackage{enumerate}
\usepackage{threeparttable}
\usepackage{makecell}
\usepackage[font=small,skip=0.2pt]{caption}
\newcommand{\subparagraph}{}
\usepackage[normalem]{ulem}
\usepackage{scalerel}
\usepackage{tikz}
\usepackage{float}
\usepackage{fontawesome}
\usetikzlibrary{calc,positioning,shapes.geometric,shapes,arrows,chains,mindmap,trees,backgrounds}

\usepackage{listings}
\usepackage{wasysym}

\ifCLASSOPTIONcompsoc
    \usepackage[caption=false, font=normalsize, labelfont=sf, textfont=sf]{subfig}
\else
\usepackage[caption=false, font=footnotesize]{subfig}
\fi

\theoremstyle{definition}

\newcommand{\cmark}{\ding{51}}%
\newcommand{\xmark}{\ding{55}}%

\definecolor{codegreen}{rgb}{0,0.6,0}
\definecolor{codegray}{rgb}{0.5,0.5,0.5}
\definecolor{codepurple}{rgb}{0.58,0,0.82}
\definecolor{backcolour}{rgb}{0.95,0.95,0.95}
\definecolor{bronze}{rgb}{0.8, 0.5, 0.2}
\definecolor{aogreen}{rgb}{0.0, 0.5, 0.0}
\definecolor{dkgreen}{rgb}{0,0.6,0}
\definecolor{gray}{rgb}{0.5,0.5,0.5}
\definecolor{mauve}{rgb}{0.58,0,0.82}
\definecolor{brilliantrose}{rgb}{1.0, 0.33, 0.64}

\theoremstyle{remark}

\makeatletter
\def\hlinew#1{\noalign{\ifnum0=`}\fi\hrule \@height #1
\futurelet\reserved@a\@xhline}
\makeatother

\definecolor{greyf}{rgb}{0.7, 0.7, 0.7}
\definecolor{greys}{rgb}{0.85, 0.85, 0.85}
\definecolor{aogreen}{rgb}{0.0, 0.5, 0.0}
\definecolor{carnelian}{rgb}{0.7, 0.11, 0.11}

\newcolumntype{a}{>{\columncolor{Gray}}c}
\newcolumntype{b}{>{\columncolor{white}}c}

\def\revise#1{{\color{black} #1}}

\newcommand{\truep}{\textcolor{aogreen}{\faCheckSquareO}}
\newcommand{\falsep}{\textcolor{carnelian}{\faTimesCircleO}}
\newcommand{\falsen}{\textcolor{carnelian}{\faCircleO}}

\newcommand{\PreserveBackslash}[1]{\let\temp=\\#1\let\\=\temp}
\newcolumntype{C}[1]{>{\PreserveBackslash\centering}p{#1}}
\newcolumntype{R}[1]{>{\PreserveBackslash\raggedleft}p{#1}}
\newcolumntype{L}[1]{>{\PreserveBackslash\raggedright}p{#1}}

\begin{document}
\title{TAPInspector: Safety and Liveness Verification of Concurrent Trigger-Action IoT Systems}
\author{Yinbo Yu, \IEEEmembership{Member, IEEE,}
        Jiajia Liu, \IEEEmembership{Senior Member, IEEE}
\thanks{This work was supported by the Basic Research Programs of Taicang (TC2020JC03), the GuangDong Basic and Applied Basic Research Foundation (2021A1515110279), the Natural Science Basic Research Program of Shaanxi (2022JQ-611), and the Fundamental Research Funds for the Central Universities (D5000210588). (Corresponding author: Jiajia Liu)}
\thanks{Y. Yu, and J. Liu are with the School of Cybersecurity, Northwestern Polytechnical University, Xi’an, Shaanxi, 710072, P.R.China (e-mail: yinboyu@nwpu.edu.cn; liujiajia@nwpu.edu.cn).}}

\maketitle

\begin{abstract}

Trigger-action programming (TAP) is a popular end-user programming framework that can simplify the Internet of Things (IoT) automation with simple trigger-action rules.
However, it also introduces new security and safety threats. A lot of advanced techniques have been proposed to address this problem. Rigorously reasoning about the security of a TAP-based IoT system requires a well-defined model and verification method both against rule semantics and physical-world features, e.g., \textit{concurrency}, \textit{rule latency}, \textit{extended action}, \textit{tardy attributes}, and \textit{connection-based rule interactions}, which has been missing until now.
By analyzing these features, we find 9 new types of rule interaction vulnerabilities and validate them on two commercial IoT platforms. We then present TAPInspector, a novel system to detect these interaction vulnerabilities in concurrent TAP-based IoT systems. It automatically extracts TAP rules from IoT apps, translates them into a hybrid model by model slicing and state compression, and performs semantic analysis and model checking with various safety and liveness properties. Our experiments corroborate that TAPInspector is practical: it identifies 533 violations related to rule interaction from 1108 real-world market IoT apps and is at least 60000 times faster than the baseline without optimization.
\end{abstract}


\IEEEpeerreviewmaketitle

\section{Introduction}

Enhanced by advanced technologies (e.g., 5G, artificial intelligence, and edging computing), the Internet of Things (IoT) promises to change the way humans interact with technology. New end-user programming frameworks also usher IoT provision into a new appified era, in which IoT services are programmable for users rather than hard-coded in proprietary equipment. Appified provision makes IoT closer to reality. Currently, \textit{trigger-action programming} (TAP) is the most popular framework for IoT programming, which enables users with or without programming experience to express their intent in automating IoT services \cite{celik2019program, wang2019charting, zhang2019autotap}. TAP has been widely supported by many commercial IoT platforms, e.g., Samsung SmartThings~\cite{smartthings}, IFTTT \cite{ifttt}, OpenHab \cite{openhab}, Mi Home, Apple HomeKit \cite{homekit}, Zapier \cite{zapier}, and so on.

With TAP, IoT systems are controlled by \textit{event-driven} apps, each \revise{consisting} of one or more TAP \textit{rules}. A TAP rule basically contains a trigger (an event or a state change) and an action (a device command) in the form of ``\textbf{IF} that \textit{trigger} occurs, \textbf{THEN} perform the \textit{action}''. Additionally, a Boolean constraint on states (\textit{i.e.}, \textit{condition}) may be also inserted in rules to enable accurate contexts. An example of TAP rules is ``\textbf{IF} the temperature drops below 10\textcelsius\ \textbf{After} 6 pm, \textbf{THEN} turn on the heater''. Given a TAP rule, an IoT system can automatically react following it without user involvement. TAP simplifies customization of IoT automation, especially for novice users.
But its limited expressivity makes users hard to construct large and intelligent automation \cite{huang2015supporting, corno2019empowering}, for which users need to design a lot of rules and coordinate them to complete tasks. \revise{The IoT physical world contains complex features (e.g., implicit dependencies between physical attributes, joint effects of device instructions). It is challenging for users to understand the physical world for expressing their intent in TAP rules, resulting} in security and safety risks due to flawed rules as well as \textit{rule interactions} \cite{wang2019charting,chi2020cross}.

Recently, many advanced techniques have been proposed to secure TAP-based IoT systems, including interaction threat detection \cite{ding2018on, wang2019charting, chi2020cross,mohannad2020scalable}, privacy leakage analysis \cite{bastys2018if, hsu2019safechain}, anomaly detection \cite{sikder20176thsense,fuhawatcher}, dynamic policy enforcement \cite{celik2019iotguard}, and automation generation \cite{zhang2019autotap, manandhar2020towards}. These techniques make \revise{significant} contributions via static analysis or formal methods. Most of them \cite{nguyen2018iotsan, chi2020cross, mohannad2020scalable, ding2018on} assume that rules are performed atomically in order and model the IoT system in a sequential model. We agree with this assumption because many triggers do not share global variables\revise{,} and a sequential model can simplify the state space of analysis. However, such \revise{a} model ignores many practical features of TAP rule executions. For instance, triggers driven by the physical world can occur concurrently\revise{,} and multiple rules can be executed in parallel \cite{mi2017an}, and thereby there may be concurrent security issues, e.g., data race. Describing an IoT system in a sequential model may introduce false results in security analysis.
Here, we scrutinize possible missed features as follows:

\textbf{Rule latency}: A TAP rule's action may not be executed immediately when its trigger activates due to delays defined in apps (e.g., \texttt{runIn} in SmartThings) or platform delays in event handling \cite{mi2017an}. Additionally, besides \textit{instantaneous} and \textit{sustained} actions \revise{that} are \revise{finished} immediately, another type of action is \textit{extended} in time (\textit{i.e.}, \revise{finished} in a period), e.g., uploading \revise{a} video. These two factors \revise{can} both delay to rule execution, \textit{i.e.}, \textit{rule latency}. For a concurrent IoT system, it is necessary to model TAP rules with rule latency, especially for detecting concurrent security issues.

\textbf{Rule interaction}: There are channels \revise{that} can chain TAP rules unexpectedly via physical attributes and underpin security issues \cite{ding2018on, wang2019charting, mohannad2020scalable}. But existing security analysis of TAP did not distinguish these attributes into immediate and tardy types, so there may be false results. This is because the \revise{latter} (e.g., temperature) typically takes some time to change to the desired value, and the time depends on the specific physical environment and device capabilities, which is different from the former (e.g., illuminance).
Besides channels, we find there is another common factor (\textit{i.e., connection}) that can cause rule interactions, e.g., \revise{the} wired connection between an outlet and its powered actuators, \revise{the} wireless connection between a ZigBee Hub and ZigBee sensors. A sound analysis of rule interactions (see $\S$\ref{sec:overview}) can ensure that our study will not miss potentially unsafe interaction behaviors.

\textbf{Safety and liveness property}: The security of TAP rules has been widely studied with various \textit{safety properties} (\textit{i.e.}, something bad should never happen). However, a richer set of properties is left out, which depends on IoT systems guaranteeing \textit{liveness} (\textit{i.e.}, something good eventually happens). They are indispensable, especially for a concurrent system \cite{owicki1982proving}. While a safety property guarantees \textit{partial correctness}, live properties correspond to \textit{total correctness}, e.g., termination without action looping, guaranteed device states. We agree that verifying both safety and liveness properties offers a more natural way to inspect TAP rules. For instance, a property ``the intrusion video is eventually uploaded to the Cloud upon physical intrusion'' contains liveness and implicit safety and can be used to check both the beginning and termination of the extended action of ``uploading \revise{a} video''.

As we know, these above features (including concurrency, rule latency, extended action, tardy-channel- and connection-based rule interaction, and liveness verification) have not yet been well studied in existing work. Hence, in this paper, we comprehensively analyze the impact of these features on TAP rule \revise{executions} and find 9 new types of rule interaction vulnerabilities. We then propose a tool named TAPInspector, based on semantic analysis and model checking to detect these vulnerabilities \revise{automatically}. To our best knowledge, TAPInspector is the first work that \revise{considers} these above features systematically for security analysis of IoT TAP rules. Our main contributions are summarized as follows:
\vspace{-0.5mm}

\begin{itemize}
  \item \textbf{9 New Types of Rule Interaction Vulnerabilities} ($\S$\ref{sec:background}-\ref{sec:overview}).
  We carefully analyze the execution process of TAP rules both in the cyber and physical world and point out essential features in physical \revise{attributes}, rule semantics, and rule latency, which require more attention in security analysis. By \revise{formalizing} the TAP-based IoT system, respecting rule semantics with rule latency and channel- and connection-based rule interactions, we uncover 9 new types of rule interaction vulnerabilities and validate them with two commercial IoT platforms.

  \item \textbf{Automatic Modeling and Verification of IoT systems} ($\S$\ref{sec:tapinspector}). By automatically extracting TAP rule semantics from source codes of IoT apps, TAPInspector constructs them with rule interactions as a finite state machine (FSM) in the concurrent format to be sensitive \revise{to} practical IoT features. To reduce the FSM's state space, TAPInspector slices it into a set of hybrid rule models and compresses variables' states in each model. In each hybrid model, rules with dependencies are executed sequentially, and rules without dependencies are executed in parallel. Finally, we design a set of safety and liveness properties according to existing and our found vulnerabilities. TAPInspector verifies rule models via rule semantic analysis and model checking.
  \item \textbf{Evaluation}($\S$\ref{sec:evaluation}). We evaluate the performance of TAPInspector for \textit{verification accuracy} with a labeled IoT app benchmark \cite{iotmal} and \textit{detection capability} with a real-world market IoT app dataset (1108 apps) from two platforms (SmartThings and IFTTT). Experimental results show that TAPInspector totally identifies 533 violations in 1108 apps and significantly reduces the verification time from minute-level to millisecond-level.
\end{itemize}

\section{Background}
\label{sec:background}
\subsection{Trigger-Action IoT Platform}
\label{subsec:tap}

To seek a rigorous security analysis of rule executions in IoT, we \revise{study} the cyber-physical gap around three nonnegligible features as follows:

\textbf{1) Attribute.} Introducing the TAP paradigm into IoT platforms provides an intuitive abstraction of IoT services and devices. We use \textit{attribute} to represent the basic abstract object that services interact with devices, including \revise{the} state of devices and system state (e.g., \texttt{locationMode}). Considering physical features, we make a distinction with two attribute types: \textbf{\textit{Immediate}}, whose value can change to another value immediately, e.g., illuminance and switch status, and \textbf{\textit{Tardy}}, whose value takes a while to change, e.g., temperature and humidity. With this distinction, we can \revise{more accurately} analyze the actual effect of rule executions.

\textbf{2) Rule Semantic.} An IoT device can be a \textit{sensor}, an \textit{actuator}, or both. A sensor measures and reports external information, e.g., a motion detector. An actuator executes device commands and manipulates corresponding attributes, e.g., a switch. By abstracting devices, the IoT platform provides a set of abstract \textit{events}, \textit{states}, and \textit{commands} for users to customize TAP rules as follows:

\begin{itemize}
  \item \textbf{Trigger}: it can be defined by an event (an instantaneous input signal, e.g., motion is detected) or a state change (a Boolean constraint on attributes becomes true, e.g., $temperature > 20\text{\textcelsius}$).
  \item \textbf{Condition}: it is a Boolean constraint defined on one or more states, and it must be evaluated to be true \revise{when rules are executed}.
  \item \textbf{Action}: two types of action (\textit{instantaneous} and \textit{sustained}) are both completed immediately and have been widely studied. While the former does not change attributes, e.g., ``send a notification'', the latter changes attributes via an actuator, e.g., ``turn on the light''. Additionally, there is a third and common type of actions, which is \textit{extended} in time \cite{huang2015supporting}. We call them \textbf{\textit{extended actions}}. An extended action usually switches the state of a device into ``doing something'' (e.g., opening, brewing) and requires a while (varying from seconds to hours) to complete. Finally, the state of the device switches to its original or another state. Examples of extended actions include ``uploading videos'' and ``turn on fan for 15 minutes''.

\end{itemize}

\textbf{3) Rule latency.} The existing literature assumes that the action of a TAP rule is executed immediately when its trigger and condition are activated. However, this is not always the case. Rule execution may involve rule latency from two aspects: \textit{trigger-to-action (T2A) latency} and \textit{delay in extended actions}. T2A latency is the delay from when the trigger occurs to when the action is executed, and can be: (1) defined in IoT apps to postpone the execution of actions within a delay, e.g., ``after a quiet period'' in IFTTT applets, \texttt{runIn(delay, method)} (actions are defined in the \texttt{method}) in SmartThings apps (\textit{i.e.}, SmartApps); (2) introduced by platform delays in event handling, e.g., long intervals in IFTTT's polling operations which are long (usually 1 to 2 minutes) with huge variance (up to 15 minutes) \cite{mi2017an}. Extended delay is the time from when an extended action starts to when it completes. Considering rule latency when inspecting threats in TAP rules enables more accurate analysis, especially for a concurrent model, but at the cost of excessive state space. This motivates us to model rules with a concurrent model and optimize the model to reduce its state space.

\subsection{Model Checking with Safety and Liveness Properties}

Model checking is a method to determine if a system satisfies a given property. A system is usually modeled as a \textit{transition system}. A property about the system is expressed in \revise{the} temporal logic formula \cite{pnueli1977temporal}, with which the model checker traverses every possible matching path of state transitions. Properties can be categorized into \textit{safety} and \textit{liveness} for a modeled system \cite{owicki1982proving, berard2001livenss}. If the system cannot meet a safety property, it implies some irremediable ``something bad'' (\textit{i.e.}, an undesired state) must happen in a path of state transitions. A liveness property asserts that it is always the case that some desired states (e.g., termination, guaranteed service) can be reached in the future, and ensures no partial path is irremediable. Liveness verification is widely applied in networks, e.g., stateful network functions \cite{farnaze2020liveness} and broadcast networks \cite{chini2019liveness}. Generally, liveness properties can be expressed in two forms \cite{owicki1982proving}: ``event A \textit{leads-to} event B'' and ``event A \textit{eventually} happens''. Some example properties that guarantee liveness in IoT systems include: 1) whenever the temperature in a room is below 18\textcelsius, setting the heater to 28\textcelsius\ \textit{leads-to} the temperature  rising to 28\textcelsius; 2) the intrusion video is eventually uploaded to the Cloud upon physical intrusion.

\section{Rule Interaction Vulnerabilities}
\label{sec:overview}

In this section, we first give our motivation and threat model. Based on our formulation of TAP rules and rule interactions, we \revise{define} 9 new types of rule interaction vulnerabilities.
\vspace{-2mm}
\subsection{Motivation and Threat Model}

\begin{figure}[!t]
\centering
  \begin{minipage}{1\columnwidth}
    \centering
    \subfloat[Rule latency]{\includegraphics[width=0.5\columnwidth]{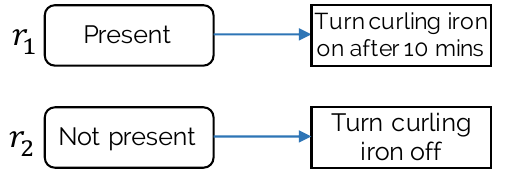}}
    \subfloat[Physical channel]{\includegraphics[width=0.5\columnwidth]{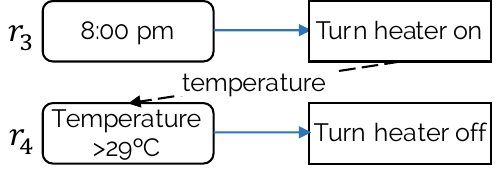}}
  \end{minipage}
  \vspace{1mm}
  \caption{Rule interactions.}
  \label{fig:channel}
  \vspace{-6mm}
\end{figure}

Rule interactions can \revise{be used to} chain rules together to provide more intelligent home automation. \revise{For example, through a temperature channel, $r_4$ in Fig. \ref{fig:channel}(b) can be activated by $r_3$.} However, most users may not realize interactions among rules in installed apps due to independent app development. Even for experienced users, \revise{it is hard to understand the IoT physical environment and various devices correctly} for app deployment and configurations \cite{zhang2019autotap}. Hence, rule interactions may often occur \revise{unexpectedly} and challenge the security and privacy of IoT systems, which have been studied in many works \cite{ding2018on,hsu2019safechain, chi2020cross,mohannad2020scalable}.

To detect vulnerabilities \revise{in rule interactions}, existing works model TAP rules into a sequential model (\textit{i.e.}, performing rules atomically in order) for model checking. However, when we consider more practical IoT features, like concurrency and rule latency, rule interactions may not act as described in such a model. Taking rules in Fig. \ref{fig:channel} for example, in a sequential model, $r_1$ and $r_2$ do not interact due to ordered executions, and $r_3$ and $r_4$ can interact through the temperature channel. Once set in a concurrent model with rule latency, we can identify a violation between $r_1$ and $r_2$ that after the user comes and leaves within 10 minutes, the curling iron will be turned on after the user leaves.
Furthermore, suppose a rule $r_5$ (``if temperature $>$ 25\textcelsius, close the outlet'') is installed with $r_3$ and $r_4$, and its outlet is incorrectly configured to the one powered the ZigBee hub of the temperature sensor, rather than the one powered the heater (\textit{i.e.}, a misconfigured physical connection), a violation that the heater cannot be turned off when the temperature exceeds 29\textcelsius\ can occur. This is because after $r_3$ is executed, $r_5$ is activated before $r_4$ due to tardy temperature and then disables the temperature sensor. $r_4$ will not be activated ever. These issues are due to the neglect of \revise{rule execution features in realistic environments.}

In this work, our goal is to detect rule interaction vulnerabilities by conducting the security analysis of TAP rules with realistic settings. We assume that TAP rules are executed in an environment \revise{that} involves more practical features, including concurrency, rule latency, extended action, distinguished physical channels (immediate and tardy), and connection-based rule interactions. We also suppose that rule interaction vulnerabilities can be induced by many methods: (1) through social engineering or phishing, an adversary can trick users into installing malicious apps, with which he can exploit weaknesses in user's configurations to generate problematic events or actions \cite{wang2019charting}; (2) an adversary can disturb the IoT system in a variety of ways, including manipulation of a 3rd party service \cite{fernandes2018decentralized}, communication failures, controlling flawed device firmware, physically spoofing sensors, \revise{etc.}; (3) non-rigorous app development leads to inconsistencies between app descriptions and code logic \cite{nguyen2018iotsan, chi2020cross}, which pose the opportunity for attackers to discover vulnerabilities; (4) complex IoT environments (e.g., concurrent rule execution and unstable rule latency) make users hard to perceive subtle rules interactions when installing apps \cite{ding2018on, wang2019charting}. A vulnerability can turn the system into an insecure or unsafe state and enable attackers to launch attacks, e.g., safety risks or privacy monitoring. Note that \revise{what kind of attack and impact vulnerabilities can be used to launch depends on the IoT device they work on, and how to launch attacks is out of our scope.}

\subsection{TAP Rule and Interaction Formulation}
\label{subsec:formulation}

Considering rule latency $l\geq 0$, we formalize a trigger-condition-action rule to be \textit{latency-sensitive} as follows:
\begin{equation}\label{equ:lrule}
    r:= (t,C_t)\overset{l}{\mapsto} (C_a,A),
\end{equation}

\noindent where $t$ is a trigger, $A$ is an action set, $C=C_t\cup C_a$ is an condition set, in which $C_t$ is a trigger condition set and is checked when $t$ is activated, and $C_a$ is an action condition set and is checked before $A$ is executed. Each component can be described by a triple $\langle\texttt{Subject: Expressions}\rangle$ as follows:

\vspace{-6mm}
\begin{eqnarray}\nonumber
  \ &\texttt{Subject} &\texttt{Expressions}\\\nonumber
  t:=&\texttt{s}^t: &\neg \texttt{a}_1\\\nonumber
  c:=&\texttt{s}^c: &(\texttt{a}_2\neq \texttt{a}_3) \wedge (\texttt{a}_4>10)\\\nonumber
  a:=&\texttt{s}^a: & \texttt{a}_5:=\texttt{v}.\nonumber
\end{eqnarray}

\noindent A subject $\texttt{s}$ can be a cyber attribute $\texttt{a}$ (e.g., time), or a device $\texttt{d}$ which is described by a set of capabilities (denoted by $\mathbb{C}^d$), each of which $\texttt{c}$ is then described by a set of attributes (denoted by $\mathbb{A}^\texttt{c}$), e.g., a temperature sensor has a capability \texttt{temperatureMeasurement} which has an attribute \texttt{temperature}. For a trigger or condition, the \texttt{Expression} is a \textit{predicate} expression (denoted by $\texttt{P}$) describing constraints to the set $\mathbb{A}$ and their compounds, which represents an instant event (e.g., doorbell ring) or a sustained state (e.g., after 5 pm). For an action, the \texttt{Expression} is an \textit{assignment} (denoted by $\texttt{Assign}$) statement to an attribute $\texttt{a}$ with a value $\texttt{v}$, which represents the command that needs the actuator to execute. Taking a TAP rule ``If the temperature drops below 10\textcelsius\ , turn on the heater'' for example, it can be formalized as $t:= tmp\_sensor:(value:=10)$ and $a:=heater\_switch:(state:=\text{ON})$. $tmp\_sensor$ and $heater\_switch$ is the formal symbol of a temperature sensor and a switch attached with a heater, respectively. $value:=10$ means the sensor's attribute $value$ is equal to 10. $state:=\text{ON}$ means the switch's attribute $state$ will be set to ``ON''.

For simplicity, we introduce a timer $\tau^r$ and divide the original $r$ containing T2A latency into two sub-rules:
\begin{align}
& r^1:=(t,C_t)\mapsto (\emptyset,\tau^r:=l),\label{eqn:r1}\\
& r^2:=(\tau^r=0,\emptyset)\mapsto(C_a,A).\label{eqn:r2}
\end{align}

\noindent $r^1$ is equipped with an action to setup a timer $\tau^r$ with a timeout value $l$. Once $\tau^r$ times out, $r^2$ performs actions of $r$. We call $r^2$ a \textit{timeout} rule. Similarly, we also use a timer $\tau^e$ and split the rule $r$ containing an extended action $a^e$ into two sub-rules:
\vspace{-3mm}
\begin{align}
& r^1:=(t,C_t)\mapsto (C_a,\{a^e,\tau^e:=l\}),\label{eqn:er1} \\
& r^2:=(\tau^e=0,\emptyset)\mapsto(\emptyset,a^{e'}),\label{eqn:er2}
\end{align}

\noindent where we use $a^{e'}$ to denote switching to another state.

Existing literature has widely studied channel-based rule interactions. Besides, we also find another type of interaction, called \textit{connection}-based interactions.
With the rule formulation, we formalize rule interactions as follows:

\textbf{Channel}-based: it can be a physical or cyber one. A \textit{physical channel} refers to a shared environment attribute $\texttt{a}$ (e.g., temperature) between actions and triggers. A \textit{cyber channel} is built on an attribute virtually defined by apps, e.g., \textit{locationMode, vacationState}.
Considering the difference between immediate and tardy channel \revise{attributes}, we define the channel-based interaction of an attribute $\texttt{a}^{ch}$ as the following:

\begin{equation}\label{equ:channel}
  \texttt{a}^{ch}: a_i(\texttt{a}^{ch}) \overset{l^{ch}}{\hookleftarrow} t_j(\texttt{a}^{ch}) \text{ or } c_j(\texttt{a}^{ch}),
\end{equation}
\noindent where $\hookleftarrow$ means a dependency relationship, \textit{i.e.}, $t_j$ or $c_j$ containing $\texttt{a}^{ch}$ can be changed by the action $a_i$; $l^{ch}$ is the delay of a tardy attribute $\texttt{a}^{ch}$ from when $a_i$ starts to when the value of $\texttt{a}^{ch}$ activates $t_j$ or $c_j$;

\textbf{Connection}-based: it works on among subjects and also has two types: physical and cyber. A \textit{physical connection} means that two devices are connected wired or wireless, e.g., a smart plug and a connected heater, a Zigbee hub and a Zigbee sensor. A \textit{cyber connection} means that an event or state change can enable or disable a rule, e.g., ``When it is after sunset, turn on a rule''. Misconfigured connections can lead to loss of safety-sensitive data or disabling security rules and devices. For instance, for a rule ``when high electric power is detected, turn off the outlet'', a user wants to activate it on an outlet connecting a heater for saving energy, but carelessly on the one connecting a water valve. If the rule is activated due to electric leakage and there is fire, the sprinkler cannot be valid since its valve cannot be opened. Formally, we define a connection-based interaction as follows:
\begin{equation}\label{equ:connection}
    \texttt{s}^{co}_p \rightarrow \texttt{s}^{co}_c: a_i(\texttt{s}^{co}_p) \hookleftarrow r_j(\texttt{s}^{co}_c),
\end{equation}
\noindent which means the validity of rules using the child subject $\texttt{s}^{co}_c$ depends on states of the parent subject $\texttt{s}^{co}_p$.

\subsection{Rule Interaction Vulnerabilities and Threats}
\label{sec:vuluerability}

\begin{figure}[t!]
\centering
  \begin{minipage}{1\columnwidth}
  \centering
    \subfloat[\textbf{V1}: Tardy-channel-based rule blocking.]{\includegraphics[width=0.45\columnwidth]{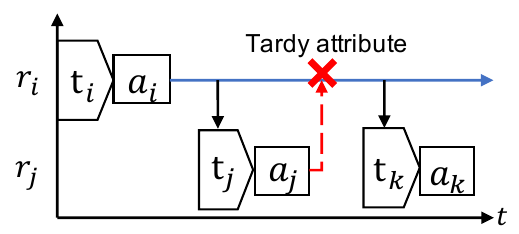}}
    \hspace{2mm}
    \subfloat[\textbf{V2}: Disordered action scheduling.]{\includegraphics[width=0.38\columnwidth]{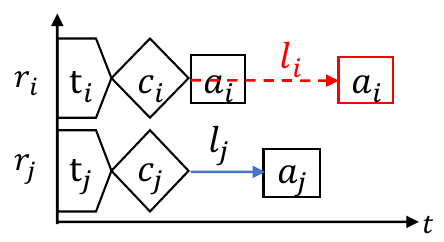}}
  \end{minipage}
    \vspace{-1mm}
  \begin{minipage}{1\columnwidth}
    \centering
    \subfloat[\textbf{V3}: Action overriding.]{\includegraphics[width=0.48\columnwidth]{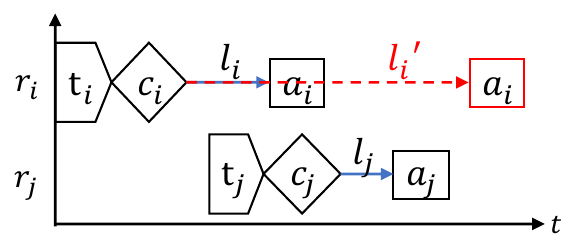}}
    \hspace{2mm}
    \subfloat[\textbf{V4}: Action breaking.]{\includegraphics[width=0.36\columnwidth]{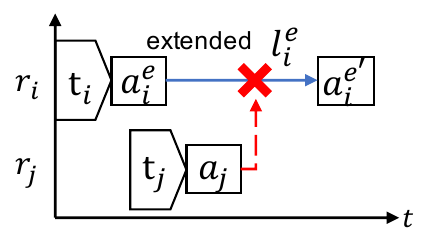}}
  \end{minipage}
    \vspace{-1mm}
  \begin{minipage}{1\columnwidth}
    \centering
    \subfloat[\textbf{V5}\&\textbf{V6}: Condition dynamic blocking.]{\includegraphics[width=0.42\columnwidth]{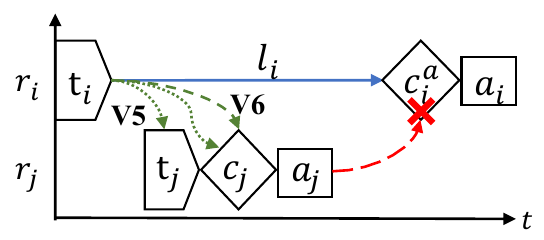}}
    \hspace{2mm}
    \subfloat[\textbf{V7}\&\textbf{V8}: Scheduled condition blocking.]{\includegraphics[width=0.4\columnwidth]{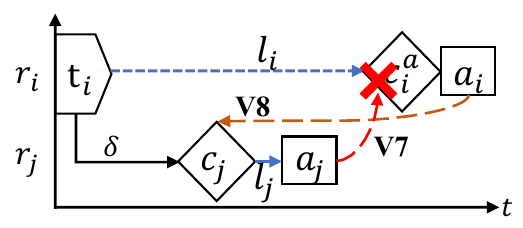}}
  \end{minipage}
  \vspace{1mm}

  \begin{minipage}{1\columnwidth}
    \centering
    \subfloat[\textbf{V9}: Device disabling.]{\includegraphics[width=0.4\columnwidth]{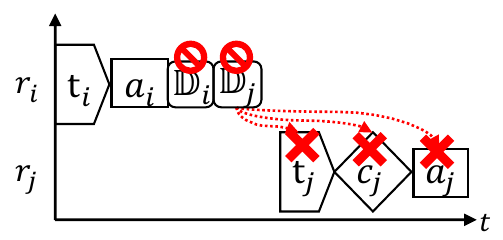}}
  \end{minipage}
  \vspace{1mm}
  \caption{Nine new types of vulnerability. We use dashed red arrows to indicate state blocking, green arrows to indicate that two constraints work on the same attribute (dotted line) or are equivalent (dashed line), dashed orange arrows to indicate condition enabling, and dotted red arrows to indicate rule disabling.}
  \label{fig:violation}
  \vspace{-4mm}
\end{figure}

Existing works \cite{wang2019charting, chi2020cross, mohannad2020scalable, nguyen2018iotsan,celik2018soteria} have defined various rule interaction vulnerabilities. \revise{We analyze the impact of different TAP rule execution features on these vulnerabilities, and uncover} 9 new types of rule interaction vulnerability summarized in Fig. \ref{fig:violation}. \revise{In this subsection, we first discuss these new vulnerabilities and then give the formula of threats for exploiting existing and our found vulnerabilities.}

\textbf{V1}: \textit{Tardy-channel-based Rule Blocking}.
\revise{Prior work \cite{ding2018on,chi2020cross, mohannad2020scalable} has studied that channel-based interactions may turn the IoT system into an unexpected state, e.g., actuation loop. Different from existing work, we distinguish immediate and tardy channel attributes and identify a new vulnerability in tardy-channel-based interactions.}
Taking a tardy attribute temperature \revise{(\texttt{Temp}) and three rules $r_i$ (``if \texttt{Temp} drops below 10\textcelsius, turn heater on''), $r_j$ (``if \texttt{Temp} exceeds \revise{20}\textcelsius, open window''), and $r_k$ (``if \texttt{Temp} exceeds \revise{26}\textcelsius, turn the heater off'') for example, the action $a_i$ (``turn heater on'') of $r_i$ can activate $r_j$ and $r_k$ through the \texttt{Temp} channel in theory. Once $a_i$ is executed, since \texttt{Temp} changes slowly, $r_j$ will be activated first to open the window which may stop \texttt{Temp} from rising and block $r_k$ to turning the heater off since the outside temperature is below 10\textcelsius.
If we do not consider the tardy change feature of attributes as prior work, vulnerability detectors will think that $r_k$ and $r_j$ are activated simultaneously, and it is a safe state. We call this vulnerability} the tardy-channel-based rule blocking shown in Fig. \ref{fig:violation}(a), in which $a_i$ can activate two rules through a tardy physical channel and \revise{one rule has an action that can change or stop the changing way of the channel.}

\textbf{V2}: \textit{Disordered Action Scheduling}.
\revise{Action conflict is a common vulnerability \cite{chi2020cross,mohannad2020scalable} in which two rules have the same trigger, mutually nonexclusive condition set, but conflicting actions. But, not all action conflicts are undesirable. There is a common setting that by designing a T2A latency, two rules can be scheduled to execute in order, similar to an extended action. For example, $r_i$: ``When motion is detected, turn on the light'') and $r_j$: ``When motion is detected, turn off the light after 1 minute''.} However, we observe that misconfigured T2A latency or unstable platform delays can disorder such scheduling and lead to violations. \revise{For example, platform delays may result in that $r_i$ has a latency greater than 1 minute and the light will not be turned off, as shown in Fig. \ref{fig:violation}(b). We call this vulnerability disordered action scheduling.}

\textbf{V3}: \textit{Action Overriding}. \revise{Similar to \textbf{V2}, we find another new vulnerability \textbf{V3} (called action overriding) regarding action conflict, but without constraints of trigger and condition. As shown in Fig. \ref{fig:violation}(c), $r_i$ is activated before $r_j$, but has a longer latency ($l_i'>l_j$) due to incorrect or varying rule latency, thereby $a_i$ will override the effect of $a_j$. Fig. \ref{fig:channel}(a) is a typical vulnerability of \textbf{V3}, in which the curling iron is turned on after the user leaves. If we ignore rule latency, we can just report action conflict, but some of which is practically correct, e.g., when $l_i<l_j$, $a_i$ and $a_j$ are executed in the original order.}

\textbf{V4}: \textit{Action Breaking}. It mainly occurs in the extended actions (see Fig. \ref{fig:violation}d), e.g., heating, uploading a video. We use the notation $a^e\overset{l^e}{\rightarrow}a^{e'}$ to represent the execution of an extended action $a^e$, \textit{i.e.}, once $a^e$ starts, the corresponding device will stay on a state for a period $l^e$ and then switches to another state via $a^{e'}$. During this period, \revise{due to event concurrency, $r_j$ may be activated with an action $a_j$ that can} stop the progress of $a_i^e$ via its action $a_j$, e.g., stopping video uploading, and result in an undesired state.

Condition blocking is a vulnerability in which \revise{an action $a_i$ can lead to a condition $c_j$ not being satisfied directly or through a physical channel} (denoted by $a_i\Rightarrow \neg c_j$) \cite{wang2019charting, chi2020cross, mohannad2020scalable}, \revise{thereby disabling} rule executions. \revise{But as studied in \cite{chi2020cross}, condition blocking may be invalid since it depends on how IoT platforms execute simultaneous rules. Hence, different from prior work, we consider the impact of rule latency, concurrency, and tardy channel for detection accuracy}, and uncover 4 new subclasses of this vulnerability: \textbf{V5}-\textbf{8}.

\textbf{V5}. As shown in Fig. \ref{fig:violation}(e), in \textbf{V5}, $t_i$ and $t_j$ or $c_j$ work on the same \revise{tardy} channel attribute but have different preferences \revise{(e.g., $t_i$: ``if the temperature exceeds 20\textcelsius'', $c_j$: ``the temperature $>$ 24\textcelsius'')}, and $a_j\Rightarrow\neg c_i^a$.
\revise{$r_i$ has a T2A latency $l_i$ that is greater than the interval ($l_{t_i\rightarrow t_j}$) for the channel attribute to rise from the value used in $t_i$ ( 20\textcelsius) to the one used in $t_j$ or $c_j$ (24\textcelsius). Hence, after two rules are both activated, $a_j$ is first executed, but $a_i$ can not be executed due to $a_j\Rightarrow\neg c_i^a$.}

\textbf{V6}. \revise{It is similar to \textbf{V5}, but in \textbf{V6}}, $r_j$ has a small or no T2A latency and the condition $c_j$ is equivalent to or satisfied after the predicate of $t_i$ is satisfied, e.g., $t_i$ (``if it is 6 pm'') and $c_j$ (``after 6:10 pm''). $t_j$ (e.g., ``user is present'') may be activated during the latency $l_i$ and can lead $c_i^a$ to be not satisfied. We call \textbf{V5} and \textbf{V6} \textit{condition dynamic blocking}. \revise{If we do not consider tardy channel and rule latency as prior work, false positive will be introduced into identified condition blocking vulnerabilities. For example, if $l_i<l_{t_i\rightarrow t_j}$, prior work still can identify condition blocking, which is actually infeasible.}

\textbf{V7}. Fig. \ref{fig:violation}(f) depicts another condition blocking vulnerability \revise{\textbf{V7}, in which, }$r_j$ is scheduled to perform at an internal $\delta$ after $r_i$ is activated, \revise{e.g., $r_i$ and $r_j$ ($\delta=$ 1 minute) discussed in \textbf{V2}. However, due to misconfigured T2A latency or long platform delays in $r_j$ (\textit{i.e.}, $l_i>\delta+l_j$),} $a_j$ is executed before $l_i$ times out and leads $c_i^a$ to be not satisfied. \revise{Only when $l_i<\delta+l_j$, such two scheduled rules can work well.}

\begin{table*}[!t]
  \centering
  \caption{Rule Interaction Threats.}
  \small
  \label{tab:threat}
  \scalebox{0.9}{
  \begin{threeparttable}
    \begin{tabular}{|l|p{3.8cm}|p{7cm}|p{4.8cm}|}
    \hline
    \textbf{ID}           & \textbf{Threat} & \textbf{Definition} & \textbf{Description} \\ \hline
    \textbf{T1}  & Action Duplication &  $\exists a_i\in A_i, a_j\in A_j, s.t.\ \textit{Sibling}(r_i,r_j)\wedge a_i=a_j$\tnote{*} &$r_i$ and $r_j$ are activated simultaneously with repeated actions. \\ \hline
    \textbf{T2}   &  Action Conflict  & $\exists a_i\in A_i, a_j\in A_j, s.t.\ \textit{Sibling}(r_i,r_j)\wedge l_i= l_j\wedge a_i=\neg a_j$ &  $r_i$ and $r_j$ are activated simultaneously with conflicting actions. \\\hline
    \textbf{T3}   &  Action-Trigger Interaction (\textbf{V1}) & $\exists a_i\in A_i, a_j\in A_j, s.t.\ \textit{Sibling}(C_i,C_j)\wedge a_i\Rightarrow t_j$\tnote{$\dag$} &  $r_i$ can activate $r_j$ directly or through physical channel. \\\hline
    \textbf{T4}  & Action Overriding (\textbf{V2-3})  &  $\exists a_i\in A_i, a_j\in A_j, s.t.\ \textit{Sibling}(C_i,C_j)\wedge l_i\neq l_j\wedge a_i=\neg a_j$& $r_j$ overrides the effect of $r_i$.\\ \hline
    \textbf{T5}  &  Action Breaking (\textbf{V4})    & $\exists a_i\in A_i, a_j\in A_j, s.t.\ a_i\overset{l^e}{\rightarrow}a_i^{\prime} \wedge a_j\Rightarrow \textit{Stop}(a_i)$\tnote{$\ddag$} &$r_j$ breaks the action in progress of $r_i$. \\ \hline
    \textbf{T6} &Condition Blocking (\textbf{V5-8}) &  $\exists c_i\in C_i^a, a_j\in A_j, s.t.\  a_j\Rightarrow \neg c_i$& $r_j$ blocks the condition of $r_i$. \\ \hline
    \textbf{T7}  &Device Disabling (\textbf{V9}) &  $\exists a_i\in A_i, s.t.\ \texttt{v}(a_i)=\texttt{off} \wedge \texttt{d}(a_i)\hookleftarrow \texttt{d}(r_j)$& $r_j$ disables the execution of $r_i$\\ \hline
    \end{tabular}
    \begin{tablenotes}
      \footnotesize
      \item[*] $\textit{Sibling}(r_i,r_j)$ means two rules have the same trigger and mutually nonexclusive condition set.
      \item[$\dag$] $\textit{Sibling}(C_i,C_j)$ means two condition set are not mutually exclusive. \qquad\qquad $\ddag$ $a_j\Rightarrow \textit{Stop}(a_i)$ means the action $a_j$ can stop the extended action $a_i$.
    \end{tablenotes}
  \end{threeparttable}}
  \vspace*{-4mm}
\end{table*}

\textbf{V8}. Still with the two scheduled rules, we define another similar vulnerability \textbf{V8} in Fig. \ref{fig:violation}(f). In \textbf{V8}, $c_j$ is enabled by $a_i$ ($a_i\Rightarrow c_j$). An example occurs in two rules $r_i$ (``If humidity drops below 38\% ($t_i$), water the garden ($a_i$)'') and $r_j$ \revise{(``If humidity drops below 38\% ($t_j$) and the watering system is turned on ($c_j$), close the watering system after 2 minutes ($\delta+l_j$)''. But due to platform delays or misconfigured T2A latency, $r_i$ has $l_i>\delta+l_j$ and can leads $c_j$ to being not satisfied after 2 minutes, }thereby the watering system will not be turned off correctly. We call \textbf{V7} and \textbf{V8} as \textit{scheduled condition blocking}.

\textbf{V9}: \textit{Device Disabling}. It is caused by incorrect device connections. As shown in Fig.
\ref{fig:violation}(g), if the parent device $\texttt{d}_i$ is closed by a rule $r_i$ (\textit{i.e.}, the value of $a_i$ is to turn $\texttt{d}_i$ off), the child device $\texttt{d}_j$ is also powered off and any rule $r_j$ depending on $\texttt{d}_j$ is then disabled. Although we do not know how different platforms actually handle offline devices, we summarize two possible settings: the platform (1) disables rules associated with offline devices; or (2) uses sensing data from their last measurement. Under the setting (1), \textbf{V9} can directly cause rule blocking. Under the setting (2), \textbf{V9} can cause rule execution to deviate from reality since the measured data for triggers \revise{may be inconsistent with the actual one.}

\revise{To exploit these above and prior found vulnerabilities in \cite{nguyen2018iotsan, wang2019charting, chi2020cross, mohannad2020scalable}, we formally define 7 \textit{rule interaction threats} in Table \ref{tab:threat}. \textbf{T1-3} and \textbf{T6} come from existing literature. \textbf{T1} and \textbf{T2} are two common threats in TAP-based IoT systems in which two rules have duplicated or conflicting actions. Action duplication is redundant (e.g., repeated notifications) or even harmful (e.g., repeated transactions). Action conflict can cause the system to enter into an unknown state depending on how the IoT platform handles the execution of simultaneous actions \cite{chi2020cross}. \textbf{T3} is a threat defined in \cite{ding2018on, mohannad2020scalable, chi2020cross}, which is related to physical channels. But they did not distinguish the effect of immediate and tardy channels and thereby introducing false positives. We aim to model such two different channels to identify vulnerabilities correctly with detailed root causes for fixing them.
\textbf{T6} is a threat defined in \cite{wang2019charting, chi2020cross}. But different from them, we take concerns on rule latency in different elements to identify our found vulnerabilities \textbf{V5-8} without false positive, which have not been studied so far. Since \textbf{V2} and \textbf{V3} has similar semantics, we formalize them into a single threat \textbf{T4}, which is similar to \textbf{T2} but has different constraints on rule latency. \textbf{T5} defines the vulnerability \textbf{V4} that can break extend actions. With the threat \textbf{T7}, we consider the vulnerability \textbf{V9} caused by mis-configured device connections.}

\vspace{-4mm}
\subsection{Vulnerability Validation}

\begin{figure}[!t]
\centering
\includegraphics[width=0.7\columnwidth]{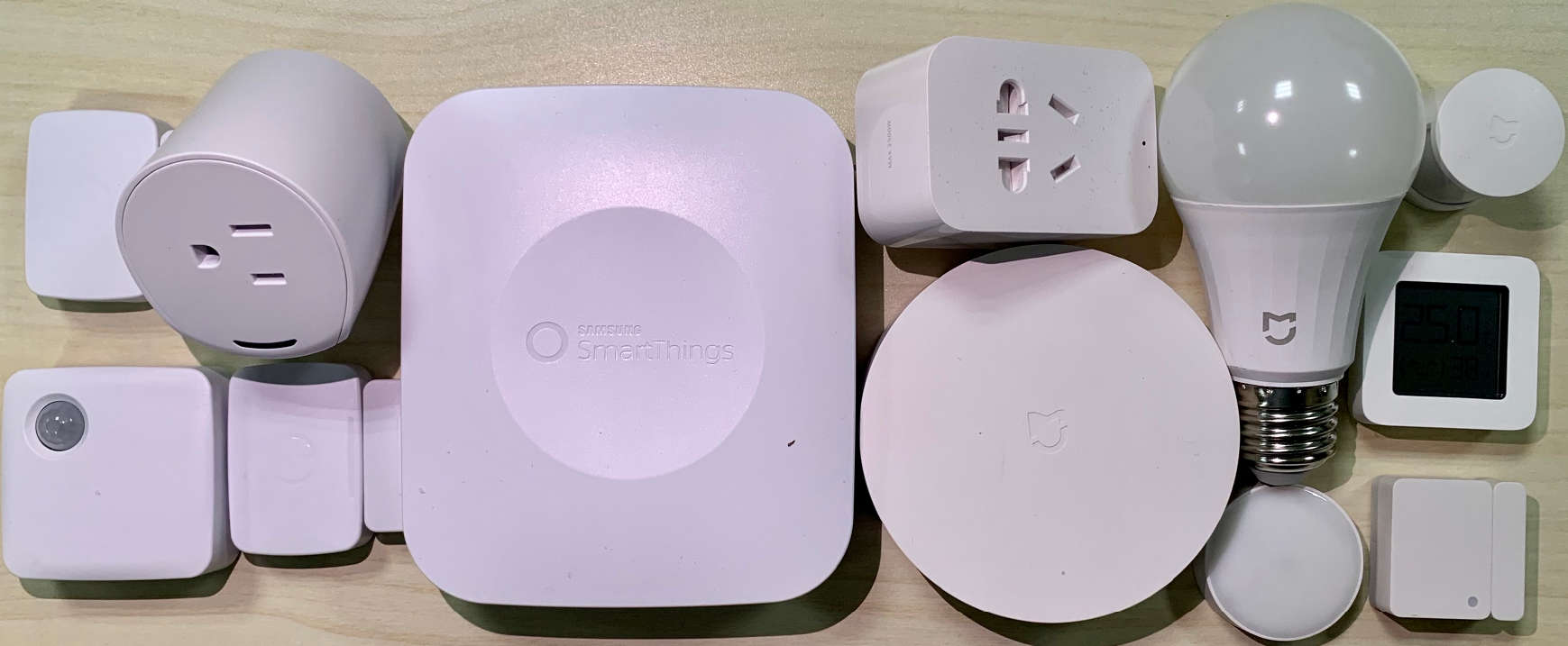}
\vspace{0.6mm}
\caption{Smart Home Devices for Vulnerability Validation.}
\label{fig:device}
\vspace{-4mm}
\end{figure}

\begin{figure}[t!]
\centering
  \begin{minipage}[c]{0.48\columnwidth}
     \subfloat[\textbf{V1}: Tardy-channel-based Rule blocking.]{\includegraphics[width=1\columnwidth]{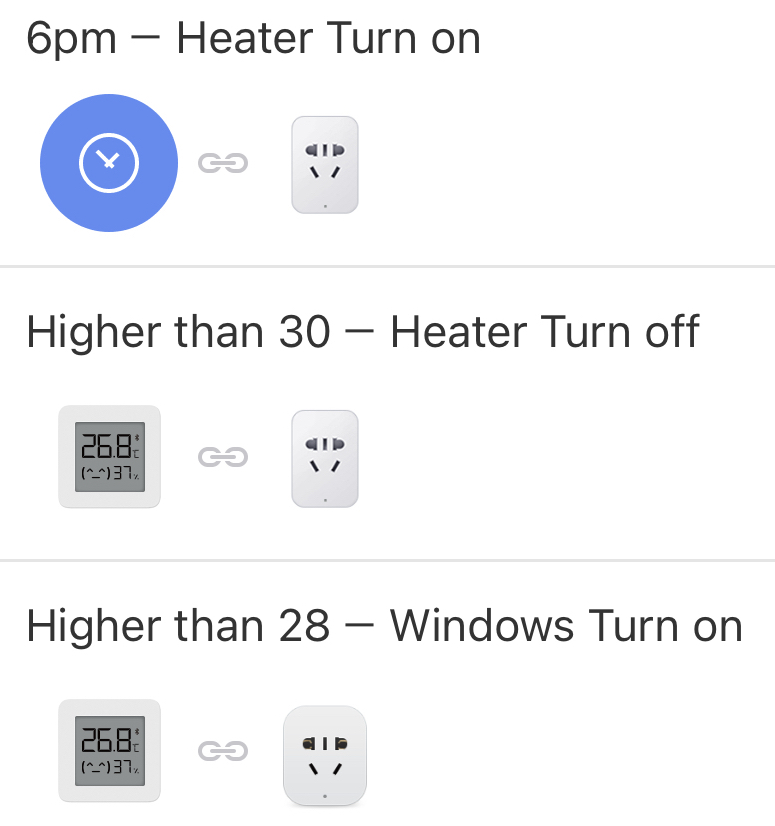}}
  \end{minipage}
  \hfill
  \begin{minipage}[c]{0.43 \columnwidth}
    \subfloat[\textbf{V2}: Disordered action scheduling.]{\includegraphics[width=1\columnwidth]{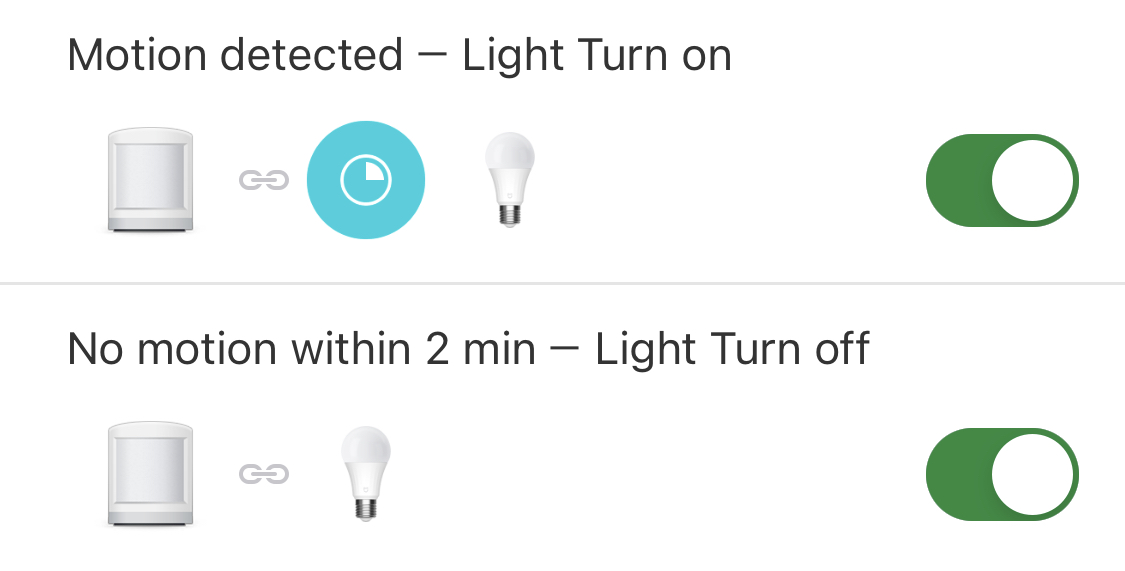}}

    \subfloat[\textbf{V3}: Action overriding.]{\includegraphics[width=1\columnwidth]{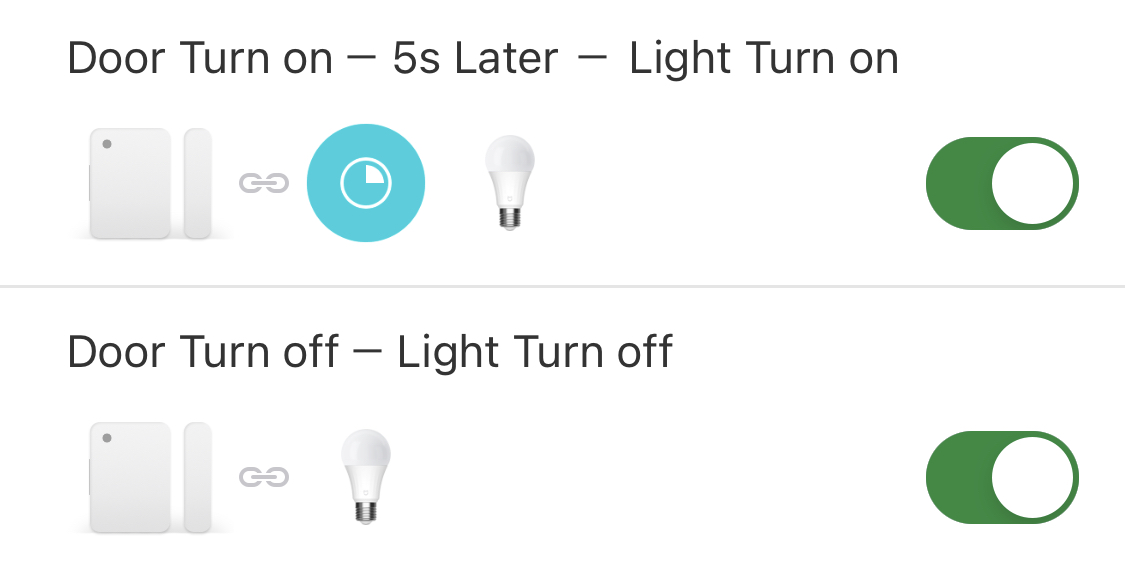}}
  \end{minipage}
  \begin{minipage}{1\columnwidth}
  \centering
    \subfloat[\textbf{V4}: Action breaking.]{\includegraphics[width=0.5\columnwidth]{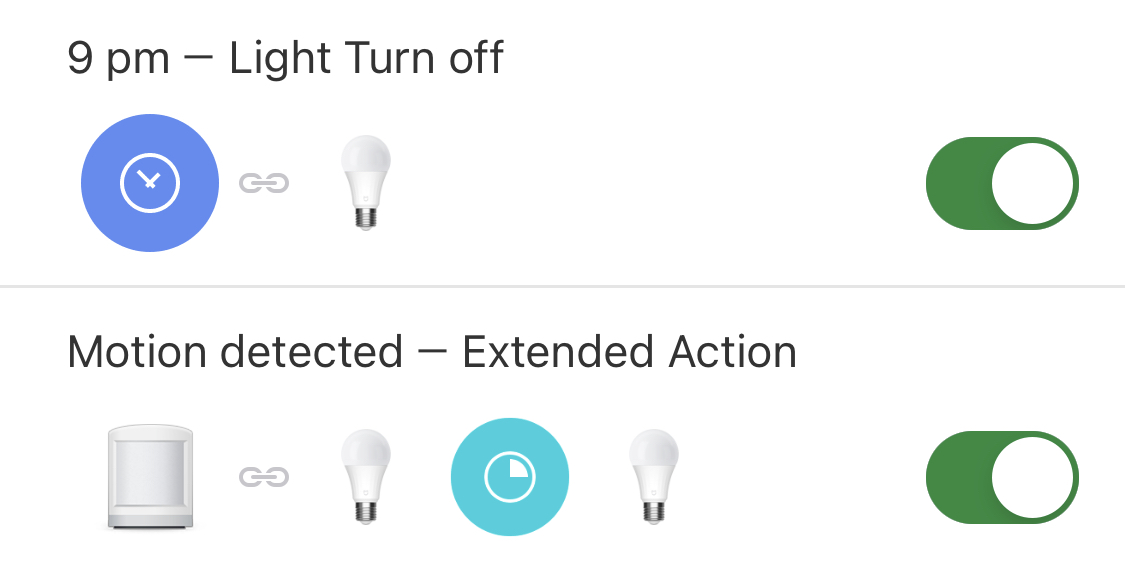}}
    \subfloat[\textbf{V9}: Device disabling.]{\includegraphics[width=0.5\columnwidth]{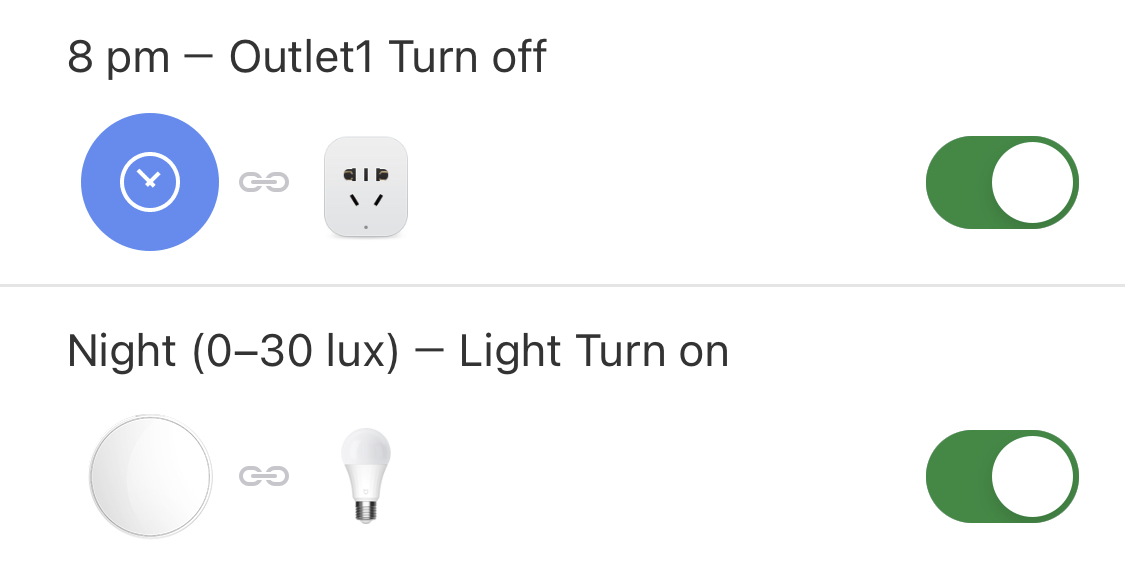}}
  \end{minipage}\vspace{1mm}
  \begin{minipage}{1\columnwidth}
  \centering
    \subfloat[\textbf{V6}: Condition dynamic blocking.]{\includegraphics[width=0.8\columnwidth]{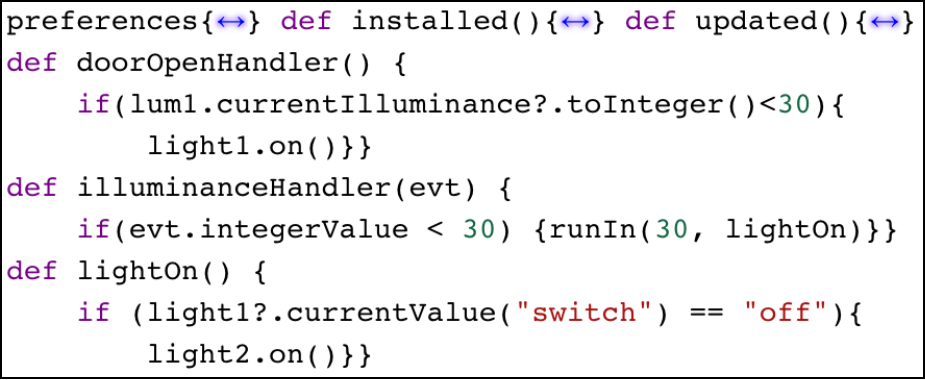}}
  \end{minipage}\vspace{1mm}
  \begin{minipage}{1\columnwidth}
  \centering
    \subfloat[\textbf{V8}: Scheduled condition blocking.]{\includegraphics[width=0.8\columnwidth]{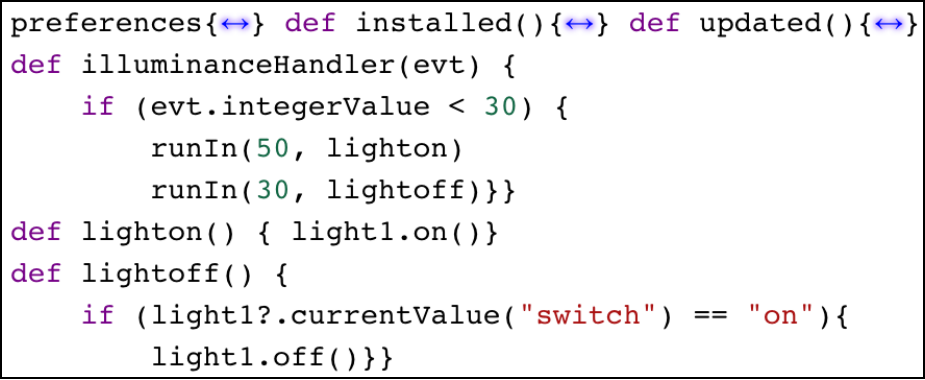}}
  \end{minipage}
  \vspace{1mm}
  \caption{TAP rules for Vulnerability Validation.}
  \label{fig:validation}
  \vspace{-6mm}
\end{figure}

Because of concurrency and unstable rule latency, our found vulnerabilities may be non-deterministic in a normal smart home setting. Hence, to validate that our found vulnerabilities are realistic and also platform-agnostic, we construct a set of TAP rules and run them on two different IoT platforms (Mi Home and SmartThings) with devices shown in Fig. \ref{fig:device}, including hub, outlet, light, motion detector, contact, luminance, and temperature sensor. TAP rules are directly set in the trigger-action form in Mi Home, called Mi automation. In SmartThings, TAP rules are coded as SmartApps in Groovy language. Mi automation and SmartApps support T2A latency via \texttt{delay} action and API \texttt{runIn(delay, method)}, respectively. We \revise{set turning a light on and off with} a delay between them \revise{as} an extended action. Fig. \ref{fig:validation} depicts our TAP rules for validating our found vulnerabilities. Fig. \ref{fig:validation} (a-e) are Mi automation we run in Mi Home with Mi Smart devices. Fig. \ref{fig:validation} (f) and (g) are two SmartApps installed in SmartThings. On \revise{these} two different commercial platforms, we can find these vulnerabilities are all valid.

\section{TAPInspector}
\label{sec:tapinspector}

\begin{figure}[!t]
\centering
\includegraphics[width=1\columnwidth]{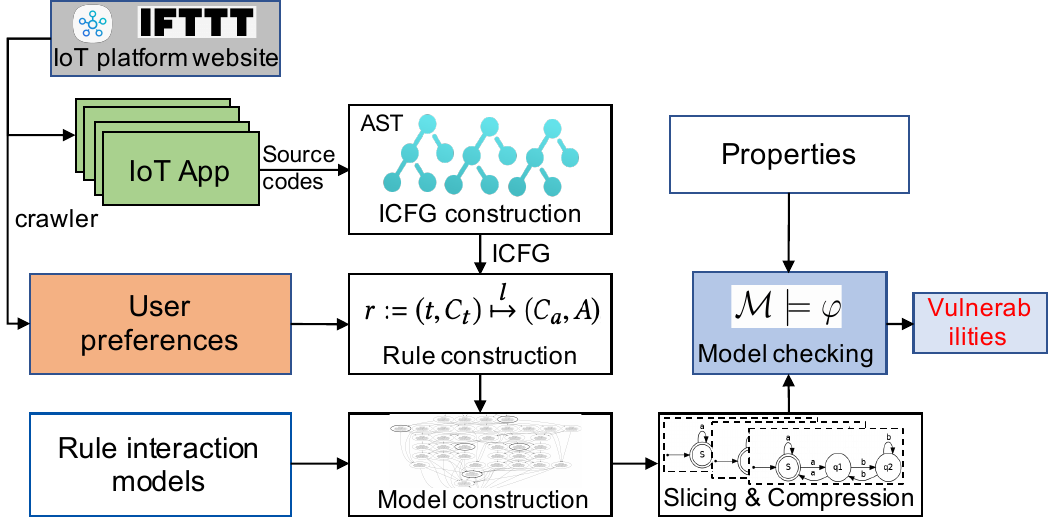}
\caption{The framework of TAPInspector.}
\label{fig:iotinspector}
\vspace{-6mm}
\end{figure}

To \revise{detect} our found rule interaction vulnerabilities in IoT apps, we present TAPInspector, a framework for automatically extracting TAP rules from IoT apps and modeling them to verify their correctness. Fig. \ref{fig:iotinspector} depicts the workflow of TAPInspector: it crawls source codes of IoT apps and corresponding user preferences from IoT platforms, extracts TAP rules from apps, constructs optimized formal models for extracted rules, and finally \revise{detects} vulnerabilities with safety and liveness properties.
In this section, we present this workflow of our TAPInspector in detail.

\vspace{-4mm}
\subsection{Rule Extraction}
The coding form of TAP rules varies from different IoT platforms. In this work, we mainly focus on SmartThings and IFTTT (rather than Mi Home) since they are the most popular IoT platforms studied in the existing literature and \revise{open-source} their market apps. Mi Home is still in the early stage, and there are only a few Mi automation public for users. Choosing SmartThings and IFTTT can also reduce the hassle of \revise{platform-specific} detection to some extent. TAPInspector first extracts inter-procedural control flow graphs (ICFGs) from IoT apps and then formalizes ICFGs as TAP rules.

\textbf{ICFG Construction}. SmartApps are coded in Groovy language and usually contain multiple rules with extensive logic. To construct the ICFG of a SmartApp, \revise{similar to \cite{nguyen2018iotsan, celik2018soteria, chi2020cross, mohannad2020scalable},} we implement an \texttt{AST visitor} (a code pass) built on Groovy compiler to traverse AST nodes at the compiler’s semantic analysis phase and perform the path-sensitive analysis \cite{dillig2008sound} to automatically extract intra-control flow graph (CFG) of each method and inter-call graph among different methods.
According to different entry methods, we classify ICFGs in two types: one is used for \textit{app initialization and schedule}, which can initialize variables both of cyber states and device attributes, subscribe event handlers via the \texttt{subscribe} method, and setup system schedules via API methods, e.g., \texttt{runIn} or \texttt{schedule}; the other type is used for \textit{event handling} declared via \texttt{subscribe}, which contains most TAP rules defined in a SmartApp. To build accurate ICFGs, we need to deal with closed-source SmartThings API methods used both for method invocations and object property access. Similar to \cite{chi2020cross}, we review the SmartThings API documentation \cite{smartapi} to model these APIs manually. For example, \texttt{runIn(delay, method)} is an API that delays the invocation of \texttt{method} within a specified \texttt{delay}. Hence, in the ICFG, TAPInspector associates the CFG of \texttt{method} with the CFG where \texttt{runIn} locates in.

IFTTT \revise{executes} rules via REST services \cite{ifttt}. Each applet typically contains a single trigger-action pair in a text-based description, e.g., ``Turn on lamp every day at 7 AM''. To \revise{formalize} IFTTT applets, NLP is a powerful but complex method \cite{wang2019charting,ding2018on}. Research works \cite{celik2019iotguard, mohannad2020scalable} present a lightweight string analysis \revise{that} tokenizes the text and \revise{searches} events and actions from tokens. Inspired by this method, we analyze the information contained in \revise{applets} crawled from the IFTTT website and find that with this information, string analysis is sufficient to convert IFTTT applets into ICFGs. \revise{Besides the text-based description, an applet also contains} several key fields describing those service APIs it used \cite{mi2017an}, including trigger channel title, trigger title, action channel title, and action title. Trigger and action channel title is the subject used in trigger and action, respectively, e.g., ``Date \& Time'' and ``WeMo Switch''. Trigger title describes an event state, e.g., ``Every data at'', ``Switched off''. Action title describes an action on the subject \revise{declared} in the action channel title, e.g., ``Turn on'', ``Start heating''. With these four fields, TAPInspector almost gets the TAP rule of an IFTTT applet. Since the trigger title may have no user configurations (e.g., ``7 AM''), TAPInspector further searches the trigger title in the applet text to identify configurations. After that, it constructs a two-node ICFG in which the entry node is the trigger, and the other node is the action. IFTTT applets use filter codes as conditions \revise{that} are not publicly visible due to privacy concerns. But we find that some rule descriptions contain a simple condition which typically is in the form of ``after a quiet period'', like ``WeMo Motion after 60min, Turn Light on''. After removing trigger and action title \revise{words} in the text, TAPInspector searches the simple condition. If found, it combines the condition into the ICFG's entry node.

\textbf{Rule Construction}. Given an ICFG, we need information about \textit{device capabilities} and \textit{user preferences} to formalize it as TAP rules. A device capability describes attributes of a device and related command methods used in apps. We extract them provided by SmartThings \cite{capabilities} and reformulate them in JSON format. Each file defines the capability of a device, including a set of attributes (each contains name, type (Boolean, Integer, or Enumerated), and a value list) and a set of command methods (each contains name and arguments). SmartThings device capabilities have covered most of \revise{the} devices used in IFTTT. For undefined devices, we manually construct them. Furthermore, SmartThings device capabilities do not distinguish \revise{between} extended action and others. Hence, we manually list a set of extended actions with compact delays. User preferences are user declarations for constants used in constraints (e.g., temperature threshold, T2A latency). User preferences are private and usually not included in source codes. To obtain them, we provide a crawler based on \texttt{Selenium} \cite{selenium} to users, with which a user can use her username and password of SmartThings to ask the crawler to extract her preferences for installed SmartApps from the platform website \cite{smartcontrol}. \revise{Besides, TAPInspector can use this crawler to collect user preferences to update the rule model and perform vulnerability detection at runtime.}

With the above information, TAPInspector extracts valid execution paths from an ICFG and formalizes \revise{them} as TAP rules as follows: for a path in \textit{app initialization}, it sets $t$ and $C$ empty and formalizes all direct or indirect (through device commands) variable-value assignment statements as the action set $A$; for a path in \textit{app schedule}, it sets up an empty rule with the action of setting up a timer of schedule delay and a timeout rule with actions $A$; and for a path in \textit{event handling}, it extracts the subject declared in \texttt{subscribe} and its associated constraints as the trigger predicate $t$, combines \revise{branch} conditions in the path into $C$, and translates commands in the path into $A$. If there is a rule latency $l$, TAPInspector reformalizes the rule as two subrules with a timer.

\subsection{Model Construction and Optimization}
\label{subsec:model}
\textbf{Model Construction}. Given a set $R$ of formulated TAP rules, TAPInspector translates them into a single finite state machine (FSM) $\mathcal{M}:=\{S,\Sigma, I\}$, where $S$ is a finite non-empty set of states, $\Sigma\subseteq S\times S$ is a state-transition function, and $I\subseteq S$ is a set of initial states. $I$ is defined by actions in those rules declared in ICFGs of app initialization and schedule. A state $s\equiv \mathbb{A}:=\mathbb{V}$ represents the attribute set $\mathbb{A}$ of all installed subjects having value $\mathbb{V}$. For other rules in $R$, TAPInspector translates each rule $r_i$ as a state-transition function $\sigma(s,s')\in \Sigma$ in $\mathcal{M}$ as follows:
\vspace{-1mm}
\begin{equation}
  s^{\prime}=\left\{\begin{array}{ll}
                    \texttt{Assign}_i(s) & \text{if } \texttt{P}^r_i(s) \wedge \texttt{P}^c_i(s)\wedge \Phi^\texttt{P}_i(s)\neq \Phi^\texttt{P}_i(s'') \\
                    s & \text{otherwise}.
                  \end{array}\right.
\end{equation}

\noindent If the current state $s$ can make the predicate of trigger and condition of rule $r_i$ satisfy ($\texttt{P}^r_i(s) \wedge \texttt{P}^c_i(s)=\text{True}$), the model performs assignments ($\texttt{Assign}_i$) of $r_i$'s actions on $s$; if not, $s$ keeps no change. For a FSM, if $\texttt{P}^r_i(s) \wedge \texttt{P}^c_i(s)$ is always true in every step, $\texttt{Assign}_i$ will be executed repeatedly, which violates real rule executions. Therefore, we introduce an auxiliary variable $\phi\in \Phi$ for each attribute to record its value in the last state $s''$ and add a new constraint $\Phi^\texttt{P}_i(s)\neq \Phi^\texttt{P}_i(s'')$ into the predicate. With this constraint, the model performs $\texttt{Assign}_i$ only when $\texttt{P}^r_i(s) \wedge \texttt{P}^c_i(s)$ becomes true.

Setting attributes of \revise{a trigger} to be non-deterministic can activate several transitions in the FSM simultaneously. \revise{Such a method can} describe the concurrency of IoT systems, but can also result in huge state space. As the number of attributes grows, the state space will grow exponentially, even explode, \revise{resulting} in a very long verification time. To address this problem, we give two optimization approaches: model slicing and state compression. The former can slice the model into a set of hybrid models, each of which has several groups executed in order, and rules in each group are executed concurrently. The latter compresses the range of variables to reduce redundant state space. 

\begin{algorithm}[!t]
  \caption{Dependency-based Model Slicing}\label{alg:slicing}
\linespread{0.98}\selectfont
\small
\DontPrintSemicolon
    \SetKwInOut{Input}{Input}\SetKwInOut{Output}{Output}
    \KwIn{An FSM model $M$ with a rule set $R$; Dependency relationships $D$}
    Collect all attributes $\mathbb{A}$  operated by all actions of $R$;\;
    Empty expression dependency edge set $E_e$;\;
    \For{$\texttt{a}_i, \texttt{a}_j\in\mathbb{A}$ and $\texttt{a}_i\neq \texttt{a}_j$}{
      Collect rule lists $R(\texttt{a}_j) \subseteq R$;\;

      \For{$r\in R(\texttt{a}_j)$}{
        \If{$\texttt{a}_i \hookleftarrow \texttt{P} \in r$}{
          add an edge $\langle\texttt{a}_i, \texttt{P} \rangle$ into $E_e$;
        }
      }
    }

    Empty rule dependency edge set $E_r$;\;
    \For{$e\in E_e$}{
      Collect rule lists $R(e.\texttt{a})\subseteq R$;\;
      \For{$r\in R$ and $\exists\texttt{P}\in r, \texttt{P}\in e$}{
        Add edges $\langle r_{s},r\rangle$ into $E_r$, $\forall r_{s} \in R(e.\texttt{a})$;\;
      }
    }

    Empty model slicer set $M$;\;
    \For{$\langle r_s, r_t\rangle \in E_r$}{
      \If{$M$ is empty or $\forall m\in M, r_s\nsubseteq m$}{
        Move $r_s$ and $r_t$ from $R$ to a new slicer $m$ in $M$;\;
        \For{$\langle r'_s, r'_t\rangle  \in E_r$}{\label{line:l2}
          \If{$r'_s\subseteq s$ and $r'_t \nsubseteq m$}{
            Move $r'_t$ from $R$ to the slicer $m$;\;
            Extract set $E'$ in $E_r$ whose source is $r'_t$\label{line:l3};\;
            \For{$r''_t \in E'$ and $r''_t \nsubseteq s$}{
            Move $r''_t$ from $R$ to the slicer $m$;\;
            \textbf{GoTo} line \ref{line:l3} with $r''_t$;\;
            }
          }\ElseIf{$r'_t\subseteq m$}{
            Move $r'_s$ from $R$ to the slicer $m$;\;
          }
        }
      }
    }
    Add remained rules in $R$ as a single slicer into $M$;\;
    \KwOut{a set of FSM model $M$}
\end{algorithm}
\setlength{\textfloatsep}{0pt}

\textbf{Model Slicing}. To slice a rule model, we enhance the model with rule interactions and design a slicing method using interaction dependencies among rules. An interaction configuration determines how an action of a device can interact with an attribute or another device, e.g., turning a heater on can increase the indoor temperature from 10\textcelsius\ to 24\textcelsius\ in 30 minutes. These configurations are obtained as follows: since cyber channels and connections are defined among rules or inside apps, we identify them by comparing all rules having operations defined in (\ref{equ:channel}) and (\ref{equ:connection}); we use physical channels identified by literature \cite{ding2018on}, but we distinguish them into tardy channels (temperature and humidity) with latency and immediate ones (illuminance, electric power, water, motion, and presence) without latency; we define our found physical connections having parent devices, including smart plug/outlet, ZigBee hub, Wi-Fi gateway, and Bluetooth hub. Child subjects are configured in different automation scenarios.

Given these channel- (\ref{equ:channel}) and connection-based (\ref{equ:connection}) interaction configurations (e.g., the list presented in Tab. \ref{tab:interconf}), we define a dependency relationship $d$ between two rules $r_i$ and $r_j$ in an FSM as follows:
\begin{equation}
  d\equiv\texttt{Assign}_i \hookleftarrow \texttt{P}_j,
\end{equation}

\noindent \textit{i.e.}, a predicate $\texttt{P}_j$ of $r_j$ depends on an assignment $\texttt{Assign}_i$ in $r_i$. Given a dependency $d$, TAPInspector inserts it into the FSM. For a channel-based dependency $d^{c}$, TAPInspector first adds the channel attribute $\texttt{a}^c$ as a variable in the FSM, and then traverses all transitions in $\Sigma$:
If a transition $r$ has an action $a$ that can change $\texttt{a}^c$ (\textit{i.e.}, $r.a\rightarrow \texttt{a}^c$) and $\texttt{a}^c$ is a tardy attribute, $\texttt{a}^c$ should be changed within several steps, rather than in one step. Hence, TAPInspector adds a new rule $r.a \overset{l^c}{\mapsto} a^c$, in which we use $r.a$ to represent a trigger when $r.a$ is executed, set a latency $l^c$ defined by $d^c$, and an action $a^c$ to change $\texttt{a}^c$;
If $\texttt{a}^c$ is an immediate attribute, we directly add the action $a^c$ into the action set of $r$;
If the trigger or condition of $r$ depends on $\texttt{a}^c$, we build a rule dependency to those having actions on $\texttt{a}^c$.
For a connection $d^{co}$, TAPInspector follows these two setting (discussed in the end of $\S$\ref{sec:vuluerability}) to insert the dependency. That is if there is a transition $r$ having a subject connected to $\texttt{s}_p^{co}$ of $d^{co}$: for setting (1), TAPInspector inserts a constraint $\texttt{s}_p^{co}=\texttt{on}$ (\textit{i.e.}, $\texttt{s}_p^{co}$ is in the state of \texttt{on}) into condition set of $r$; for setting (2), TAPInspector inserts this constraint into the transition of obtaining sensing data.

After enhancing the model, TAPInspector performs model slicing \revise{following with} Algorithm \ref{alg:slicing} in three stages: (1) \textit{building expression dependency edges}: TAPInspector collects all used attributes and figures out the expression dependency set $E_e$ (Line 2-7). An expression dependency means that for two attributes $\texttt{a}_i$ and $\texttt{a}_j$, a rule uses $\texttt{a}_j$ and also has a constraint $\texttt{P}$ depending on $\texttt{a}_i$; (2) \textit{building rule dependency edges}: TAPInspector traverses the expression dependency set $E_e$ to construct rule dependency edge set $E_r$ (Line 9-12). A rule dependency edge means if a rule $r$ uses the source attribute $e.\texttt{a}$ of an expression dependency $e\in E_e$ and contains a sink $\texttt{P}$ in $e$, all rules $r_s$ using $e.\texttt{a}$ have the rule dependency on $r$; (3) \textit{slicing model}: TAPInspector performs forward traversal on the model using $E_r$ to slice the model $M$ into a set of FSM model (\textit{i.e.}, slicers) $M$ (Line 13-26). If for two rule dependency $e$ and $e'$, a slicer $m$ contains $e$ and $r'_s\in e'$ without $r'_t\in e'$, TAPInspector performs forward traversal of the model along with sink rules of $r'_t$. If $r'_t$ is contained in $m$, TAPInspector adds $r'_s$ into $m$. Finally, TAPInspector adds remained rules as a single slicer into $M$ and outputs a set of hybrid FSMs. Such hybrid models ensure that our analysis can cover both concurrency- and sequency-related features of TAP rule executions.

\begin{table*}[t!]
  \centering
  \caption{Safety and liveness properties.}
  \label{tab:property}
  \small
  \scalebox{0.78}{
\begin{threeparttable}[b]
  \begin{tabular}{|c|l|}
  \hline
  Property       & Description \\ \hline
  \textbf{T1}\tnote{*} & 1) $a_i$ leads-to $a_j$ via a tardy channel attribute; 2) When $a_i$ is executed, $a_j$ should be executed via a immediate channel attribute \\
  \hline
  \textbf{T2} & N/A since two rules have the same semantics \\
  \hline
  \textbf{T3} & When $t_i$ and $C_i$ is activated, $a_i$ should be executed\\
  \hline
  \textbf{T4} & 1) When $t_i$ is activated, $a_j$ should not be executed; 2) $t_i$ and $C_i$ leads to $a_i$ \\
  \hline
  \textbf{T5} & 1) $a_i$ leads-to $a_i^{\prime}$; 2) before $l^e$ times out, $a_i^{\prime}$ should not be executed\\
  \hline
  \textbf{T6} & When $t_i$ is activated, $a_i$ should be executed\\
  \hline
  \textbf{T7} & If $\texttt{v}(a_i)=\texttt{off}$, $a_j$ can execute \\
  \hline  \hline
  S.1 & When no one is present, the curling iron should not be turned on\\\hline
  S.2 & When the door is closed, the bell should not chime \\\hline
  S.3 & When motion is detected, the siren should be activated \\\hline
  S.4 & When the electric power is greater than a predefined value\tnote{$\dag$}, the oven and heater should not be turned on at the same time\\\hline
  S.5 & When no one is present, \texttt{locationMode} should be Away \\\hline
  S.6 & When someone is present, \texttt{locationMode} should be Home \\\hline
  S.7 & When the user is not present or sleeping, the door should not be unlocked\\\hline
  S.8 & The light must be on when the user arrives home if it is night\\\hline
  S.9 & The AC and heater must not be on at the same time\\\hline
  S.10 & When smoke is detected, the water valve should not be closed \\\hline
  S.11 & When the user is not at home, the security system should not be disarmed \\\hline
  S.12 & When a camera does not recognize an unauthorized face, the door should not be unlocked\\\hline
  S.13 & When the moisture sensor detects a leak, the water valve should be closed \\\hline
  S.14 & When the heater is on, the windows should not be open\\\hline
  S.15 & When the user is not present, the brewing functionality of coffee machine should not be activated \\\hline
  S.16 & When the user is not present, the heating functionality of electronic blanket should not be activated\\\hline
  S.17 & When the smoke is detected, the alarm should sound\\\hline
  S.18 & When the user arrives home, the garage door should be open\\\hline
  S.19 & When the user leaves home, the garage door should be closed\\\hline
  S.20 & When \texttt{locationMode} is Away, the light should not be on\\\hline
  S.21 & When no introduer is detected, the siren should not be activated \\\hline
  S.22 & When an introduer is detected, the siren should be activated\\\hline
  S.23 & At everyday noon, the water temperature should not be below a predefined value\\\hline
  S.24 & When no one is at home, the oven should not be on the status of grilling\\\hline
  S.25 & When the energy price is greater then a predefined level\tnote{$\dag$}, some plugs should not be turned on \\\hline
  S.26 & When CO2 is greater than 1000ppm, the fan should be turned on \\\hline
  S.27 & When CO is greater than a predefined level\tnote{$\dag$}, the alarm should be turned on\\\hline
  S.28 & When CO is greater than a predefined level\tnote{$\dag$}, the natural gas hot water heater should not be turned on \\\hline
  S.29 & When CO is greater than a predefined level\tnote{$\dag$}, the gas valve should not be turned on \\\hline
  S.30 & When the temperature is below a predefined value\tnote{$\dag$}\ and no one is present, the heater should not be turned on \\\hline
  S.31 & When the temperature is below a predefined value\tnote{$\dag$}\ and someone is present, the heater should be turned on\\\hline
  S.32 & When the temperature exceeds a predefined value\tnote{$\dag$}, the heater should not be turned on\\\hline
  S.33 & When the humidity is greater than a predefined value\tnote{$\dag$}, the humidifier should not be turned on\\\hline
  S.34 & When no one is at home, the electric blanket should not be turned on\\\hline
  S.35 & When there is a motion, the security camera should be turned on \\\hline
  S.36 & When the indoor illuminance is less than 80lux in daytime, the light should not be turned off\\\hline
  S.37 & When the indoor illuminance is greater than 600lux, the light should not be turned on\\\hline
  L.1 & The camera video is eventually uploaded to the Cloud after starting to upload \\\hline
  L.2 & Turning on AC leads to temperature drawing within a predefined range\tnote{$\dag$} \\\hline
  L.3 & The presence of users leads to humidity keeping within a predefined range\tnote{$\dag$} \\\hline
  L.4 & After the oven starts, it will eventually finish cooking \\\hline
  L.5 & After 10 pm, the light will eventually be turned off \\\hline
  L.6 & The coffee maker can eventually finish coffee after it starts brewing coffe \\\hline
  L.7 & After the water heater is turned on, the water temperture will eventually reach to a predefined value\tnote{$\dag$}\\\hline
  L.8 & After indoor temperature is less than 8\textcelsius, it will eventually exceed a predefined value\tnote{$\dag$}\\\hline
  L.9 & After indoor temperature is greater than 30\textcelsius, it will eventually drop below a predefined value\tnote{$\dag$} \\\hline
  L.10 & After CO2 is greater than 1000ppm, it will eventually drop below 450ppm\\\hline
  L.11 & After ``turn on fan for 15 minutes'' is on, the fan will be eventually  turned off \\\hline
  L.12 & After turning on the humidifier, the indoor humidity will keep within a predefined range\tnote{$\dag$}\\\hline
  L.13 & Once the heating set point of the thermostat is changed to 25\textcelsius, the indoor temperature will be eventually greater than 23\textcelsius \\\hline
  L.14 & Once the cooling set point of the thermostat is changed to 20\textcelsius, the indoor temperature will be eventually less than 22\textcelsius \\\hline
  L.15 & After boosting AC to a defined value\tnote{$\dag$}, the indoor temperature will eventually reach to the value \\\hline
  L.16 & After boosting the water heater, the water temperature will be eventually  greater than 60\textcelsius \\\hline
  L.17 & After the user wakes up, he will eventually get a cup of hot coffee\\\hline
  L.18 & After the user leaves, the temperature will eventually drop below \\\hline
  L.19 & Motion detected can lead the light to blink\\\hline
  L.20 & When the user pins a photo, the light's color will eventually change to match the photo \\\hline
  L.21 & After the robot cleaner starts, it will eventually come back to charge \\\hline
\end{tabular}
\begin{tablenotes}
  \footnotesize
   \item[*] While properties of \textbf{T3-7} are used to find conforming counterexamples, the properties of \textbf{T1} are used to detect the absence of \textbf{T1}.
   \item[$\dag$] TAPInspector fills them according to user preferences or randomly.
 \end{tablenotes}
\end{threeparttable}
}
\vspace{-4mm}
\end{table*}

\textbf{State Compression}. There are many integer variables representing attributes in the IoT system, such as time and illuminance. They may have a huge range, e.g., the indoor illuminance can be 0 to 2000 $lux$, \revise{leading} to a large and redundant state space for model checking. Hence, we optimize these variables as integers within a compact range in the FSM to avoid state explosion.
Specifically, given an FSM, TAPInspector lists all used values of an integer variable, maps them into a small integer range, and re-declares the variable with this range.
For instance, we use time points involved in all rules to form a range for the cyber time, rather than increase the time as a continuous counter in each transition.

While compressed variables are meaningful for immediate attributes, \revise{they} can raise new issues for tardy attributes, especially with concurrent events. For example, there are three rules: $r_1$ (``after 8pm, turn on the thermostat to 28\textcelsius''), $r_2$ (``If the temperature exceeds 25\textcelsius, turn off the thermostat''), and $r_3$ (``If the temperature exceeds 27\textcelsius, open the window''). If we simply compress the temperature as an integer in the range of [0,3] and set it directly change to 3 (28\textcelsius) from 0 in a few transitions after $r_1$ is executed, $r_2$ and $r_3$ will be executed simultaneously, which violates the fact. Hence, after listing all involved values of a tardy attribute, we add transitions of value changes to cover all of them, e.g., 1 (25\textcelsius) and 2 (27\textcelsius).

\subsection{Vulnerability Detection}
\label{subsec:modelchecking}

Given an FSM of IoT systems, TAPInspector detects \revise{our defined and existing rule interaction vulnerabilities} both by rule semantic analysis and model checking with a set of security properties. With threat formulas in Table \ref{tab:threat}, TAPInspector first compares each rule action with all components of another rule to check for interaction threats. Once a threat is identified, to avoid false \revise{positives}, TAPInspector generates a safety or liveness property according to its formula (shown in Table \ref{tab:property}) and validates the threat via model checking. For instance, an action breaking threat (\textbf{T5}) is found by rule comparison in two rules ``When the user arrives at home ($t_1$), brew coffee ($a_1$)'' and ``When sunset ($t_2$), close the outlet $a_2$'', in which \revise{the outlet powers the coffee maker}. Then, TAPInspector generates a liveness property ``$a_1$ leads-to finishing brewing coffee ($a_1^{\prime}$)'' for model checking. Finally, a violation is confirmed that the user arrives at home a short time before sunset, the coffee cannot be finished.

Additionally, we further design a set of specific safety and liveness properties, which can be expressed \revise{in} \textit{Linear Temporal Logic} (LTL) or \textit{Computing Tree Logic} (CTL) \cite{pnueli1977temporal}. Table \ref{tab:property} lists all safety and liveness properties we designed for model checking. For safety properties, we define undesirable behaviors used in prior works \cite{nguyen2018iotsan, celik2018soteria, zhang2019autotap, mohannad2020scalable}. For example, the property S.1 can be expressed as $\textbf{G}(\texttt{user}\neq present\wedge \texttt{curling-iron}:=\text{ON})$ (\textbf{G} is the ``always Globally'' LTL operator) and is used to ensure the curling iron should always be turned off when the user is not at home.
Besides, we further carefully define a set of liveness properties, especially for termination of extended actions and guaranteed states. For instance, L.1 can be expressed as $\textbf{AG}(\texttt{video}:=uploading\rightarrow \textbf{AF}(\texttt{video}:=uploaded))$ and is used to ensure that the video being uploaded can eventually be uploaded. \revise{These liveness properties in Table II are assumptions based on device behavior, physical attribute, and system state.}
In total, we consider 67 properties, including 37 safety, 21 liveness, and 9 generals of rule interaction threats.

\section{Implementation and Evaluation}
\label{sec:evaluation}
\subsection{Implementation}
To evaluate TAPInspector, we implement a prototype system of TAPInspector in Java with over 9K lines of code, including three parts: (1) \textit{ICFG Extraction}, including a Groovy AST visitor for SmartApps and a string analyzer for IFTTT applets, (2) \textit{Rule Construction}, and (3) \textit{Model Construction and Optimization}. Besides, TAPInspector uses NuSMV \cite{alessandro2002nusmv}, a mature symbolic model checker, for model checking. To perform model checking using NuSMV, TAPInspector translates FSMs into SMV files containing three modules (\texttt{CLOCK}, \texttt{TIMER}, and \texttt{main}) for the process of the cyber time, timer used in rule latency, and TAP rule enforcement, respectively. With a temporal property, NuSMV confirms it holds or not with a concrete reason by providing a counterexample.

Additionally, to enable the rule model to perceive physics, TAPInspector needs to know rule interaction configurations. For channel-based interactions, \revise{we need} to know which device actions can change a channel attribute and then how long it will take to change the attribute (especially for tardy attributes) to a defined value. Hence, we survey the effect of different IoT devices and manually construct a set of configurations for different physical channel-based interactions. In Table \ref{tab:interconf}, we show partial configurations of physical channel-based interaction. For an immediate attribute, we set which devices change it and what value the device will change to. For example, turning a light \revise{at night} on can change the illuminance to 200 lux immediately.  For a tardy channel, besides the same configuration, we also configure how much time a device needs to change from one value to another value. For example, we assume that setting the air conditioner to cooling mode can change the indoor temperature to 18\textcelsius\ in 30 minutes. Other channel configurations are constructed \revise{similarly}. For connection-based interactions, we do not inject them into the rule model directly, but just randomly pick up some pairs of a switch-capable device and other devices to set connection-based dependencies in related TAP rules. They simulates misconfigured connections.

\begin{table}
  \caption{Partial Configurations of Channel-based Interaction.}
  \label{tab:interconf}
  \scalebox{0.88}{
  \begin{tabular}{|c|c|l|}
    \hline
    Attribute & Action & Effect \\\hline
    \multirow{8}{*}{Temperature} & \texttt{ACMode.cool} & $\texttt{Temp}_{in}\rightarrow$ 18\textcelsius\ in 30 min\\\cline{2-3}
& \texttt{ACMode.heat} & $\texttt{Temp}_{in}\rightarrow$  25\textcelsius\ in 30 min\\\cline{2-3}
& \texttt{ACMode.off} & $\texttt{Temp}_{in}\rightarrow\texttt{Temp}_{out}$ in 20 min\\\cline{2-3}
& \texttt{thermostatMode.heat} & $\texttt{Temp}_{in}\rightarrow$ 18\textcelsius\ in 30 min\\\cline{2-3}
& \texttt{thermostatMode.cool} & $\texttt{Temp}_{in}\rightarrow$ 25\textcelsius\ in 30 min\\\cline{2-3}
& \texttt{window.open} & $\texttt{Temp}_{in}\rightarrow \texttt{Temp}_{out}$ in 10 min\\\cline{2-3}
& \texttt{heater.on} & $\texttt{Temp}_{in}\rightarrow 25$\textcelsius\ in 10 min\\\cline{2-3}
& \texttt{heater.off} & $\texttt{Temp}_{in}\rightarrow \texttt{Temp}_{out}$ in 30 min\\\hline
\multirow{2}{*}{Humidity} & \texttt{humidifier.on} & $\texttt{Humid}_{in}\rightarrow 45$ in 20 min\\\cline{2-3}
& \texttt{humidifier.off} & $\texttt{Humid}_{in}\rightarrow\texttt{Humid}_{out}$ in 50 min\\\hline
\multirow{2}{*}{Illuminance} &\texttt{light.on} at night & $\texttt{Illum}_{in}\rightarrow 200 lux$\\\cline{2-3}
&\texttt{light.off} at night & $\texttt{Illum}_{in}\rightarrow\texttt{Illum}_{out}$\\\hline
\multirow{4}{*}{Power} &\texttt{light.on} & $\texttt{P}+50W$\\\cline{2-3}
&\texttt{heater.on} & $\texttt{P}+2200W$\\\cline{2-3}
&\texttt{water\_heater.on} & $\texttt{P}+1500W$\\\cline{2-3}
&\texttt{AC.on} & $\texttt{P}+2800W$\\\hline
  \end{tabular}
  }
\end{table}

\subsection{Evaluation Setup}

With our TAPInspector prototype, we perform evaluations from three aspects: detection accuracy, vulnerability detection of real-world IoT apps, and performance benchmarks. The evaluations of detection accuracy are used to validate TAPInspector's accuracy for detecting different property violations. We use a public benchmark, IoTBench \cite{iotmal}, which provides flawed SmartApps that contain various property violations in a single SmartApp or a group of SmartApps \cite{celik2018soteria} and has been used as ground truth of violation detection in several literature \cite{celik2018soteria, nguyen2018iotsan, mohannad2020scalable}. Then, we use a set of market IoT apps to evaluate if TAPInspector can identify violations in practice and how many violations it can identify. We gather 1108 market IoT apps from two sources: (1) we obtain 71 SmartApps in SmartThings public repository \cite{repository} after removing those SmartApps using HTTP requests; (2) we obtain 1037 IFTTT applets from the IFTTT dataset used in \cite{mi2017an}, whose elements are defined by those device capabilities provided in \cite{capabilities}. Finally, we conduct the evaluations of TAPInspector's performance for rule extraction and formal verification of IoT apps. All of them are conducted on a desktop with 3GHz Intel Core i7-9700 and 16GB memory.

\subsection{Accuracy Validation}
\label{sec:accuracy}

To validate the accuracy of TAPInspector, we perform a fair and effective comparison with the state-of-the-art on IoTBench. However, we find that most of existing approaches \cite{chi2020cross,celik2019iotguard,ding2018on,bastys2018if} are not open-source, excluding IoTSAN \cite{nguyen2018iotsan} and \textsc{IotCom} \cite{mohannad2020scalable}. Hence, we only focus on SOTERIA \cite{celik2018soteria}, IoTSAN, and \textsc{IotCom}, whose paper or technical report \cite{celik2018soteriat} provides the results of inspecting SmartApps in IoTBench. SmartApps \textit{ID 1-9} in IoTBench contain different vulnerabilities of action duplication and action conflict. \textit{Group 1-3} in IoTBench contain violations caused by rule interactions through cyber channels. SmartApps in IoTBench do not involve violations related to physical channel, concurrency, rule latency, and physical connection. We include 3 SmartApp groups (\textit{Group 4-6}) designed via \textsc{IotCom}, which involve violations related to physical channels.

\begin{table}[!t]
    \caption{Text description of new designed malicious SmartApps.}
    \label{tab:maliciousapp}
    \centering
    \scalebox{0.95}{
    \begin{threeparttable}[b]
    \begin{tabular}{|c|p{0.9cm}|p{6cm}|}
        \hline
        \multirow{2}{*}{Benchmark} & Vulner- abilities & \multirow{2}{*}{Text-based description}\\\hline
         \multirow{2}{*}{\textit{N1}} & \multirow{2}{*}{\textbf{V3}} &If the user present, turn curling iron on after 10 minutes; If not present, turn curling iron off.\\\hline
         \multirow{2}{*}{\textit{N2}} & \multirow{2}{*}{\textbf{V2}} & Unlock and then lock the door if user present; Lock the door if not present. \\\hline
         \multirow{2}{*}{\textit{N3}} & \multirow{2}{*}{\textbf{V4}} & Brewing coffee if user is present; Close coffer maker if it is sleep time.\\\hline
         \multirow{4}{*}{\textit{Group N4}\tnote{*}} & \multirow{4}{*}{\textbf{V9}(\textbf{V1})} & If temperature drops below 20\textcelsius, turn heater on; If temperature rises above 30\textcelsius, turn heater off; If electric power rises above 3000W for 10 minutes, turn outlet off.
         \\\hline
        \multirow{6}{*}{\textit{Group N5}} & \multirow{6}{*}{\textbf{V5}} & When temperature rises above 26\textcelsius, turn heater off after 20 minutes if user is present and window is closed; If user present and temperature is below 18\textcelsius, turn heater on and close window; If temperature rises above 28\textcelsius, open window; If temperature drops below 15 \textcelsius, close window.
         \\\hline
    \end{tabular}
    \begin{tablenotes}
    \footnotesize
     \item[*] The outlet is misconfigured to the one attached the Bluetooth hub for connecting the temperature sensor, rather than the one attached the heater.
   \end{tablenotes}
    \end{threeparttable}
    }
    \vspace{-5mm}
\end{table}
\setlength{\textfloatsep}{0pt}

\begin{table}[t!]
\caption{Accuracy comparison of violation detection between SOTERIA \cite{celik2018soteria}, IoTSAN \cite{nguyen2018iotsan}, \textsc{IotCom} \cite{mohannad2020scalable}, and TAPInspector. We use \truep, \falsen, and \falsep\ to denote true positive, false negative, and false positive, respectively.}
\label{tab:accuarcy}
\small
\centering
\begin{threeparttable}[b]
  \begin{tabular}{l||c|c|c|c}
      \hline
      Benchmark & SOTERIA& IoTSAN  & \textsc{IotCom}  & TAPInspector \\
      \hline
      \multicolumn{5}{c}{\textbf{Individual App}}\\\hline
      ID 1 & \truep & \truep& \truep& \truep \\\hline
      ID 2& \falsep & \falsen& \truep& \truep \\\hline
      ID 3& \truep & \falsen& \truep& \truep \\\hline
      ID 4\tnote{*}& \truep\falsen & \falsen& \truep& \truep \\\hline
      ID 5.1& \falsen & \falsen& \falsen& \truep \\\hline
      ID 6& \truep & \truep& \truep& \truep \\\hline
      ID 7& \truep & \falsen& \truep& \truep \\\hline
      ID 8& \truep & \truep& \truep& \truep \\\hline
      ID 9& \falsep & \falsen& \truep& \truep \\\hline
      N1 & \falsen & \falsen& \falsen& \truep \\\hline
      N2 & \falsen & \falsep& \falsep& \truep \\\hline
      N3 & \falsen & \falsen& \falsen & \truep \\\hline
      \multicolumn{5}{c}{\textbf{App Group}}\\\hline
      Group 1 & \truep & \truep& \truep& \truep \\\hline
      Group 2 & \truep & \falsen& \truep& \truep \\\hline
      Group 3 & \truep & \falsen& \truep& \truep \\\hline
      Group 4 & \falsen & \falsen& \truep& \truep \\\hline
      Group 5 & \falsen & \falsen& \truep& \truep \\\hline
      Group 6 & \falsen & \falsen& \truep& \truep \\\hline
      Group N4 & \falsen & \falsen& \falsen& \truep \\\hline
      Group N5 & \falsen & \falsen& \falsen & \truep \\\hline
  \end{tabular}
  \begin{tablenotes}
    \footnotesize
     \item[*] ID 4 contains two violations: action duplication and action conflict. SOTERIA can identify only repeated actions, while \textsc{IotCom} and TAPInspector can identify both.
   \end{tablenotes}
\end{threeparttable}
\end{table}

Besides, we further develop 3 individual SmartApps (\textit{N1-3}) and 2 SmartApp groups (\textit{Group N4-5}) with designed violations related to concurrency, rule latency, extended action, and connections. For saving space, we give a text-based description of these SmartApps in Table \ref{tab:maliciousapp}. \textit{N1} contains two rules shown in Fig. \ref{fig:channel}(a) and involves a vulnerability \textbf{V3}. Hence, we can check \textit{N1} with the property S.1 in Table \ref{tab:property} for threat \textbf{T4}. In \textit{N2}, we randomly set the latency into the two actions of ``unlock the door'' and ``lock the door'' of the first rule to simulate random platform delays. Such random latency can cause a vulnerability \textbf{V2} in which the scheduled sequence of unlocking and locking is disordered. \textit{N3} contains an extended action (brewing coffee) and a time-related event (sleep time) that can stop the extended action. Hence, there is a vulnerability \textbf{V4} in \textit{N3}. \textit{Group N4} contains 3 SmartApps interacted both with a physical channel (temperature between the heater and temperature sensor) and a misconfigured physical connection in which the controlled smart plug is misconfigured as the plug attached \revise{to} the Bluetooth hub. Hence, \textit{Group N4} involves a violation \textbf{V9} which disables the temperature sensor and leads the heater to be not turned off when the temperature rises above 30\textcelsius. This violation also conforms to \textbf{V1} since rising the temperature from 20\textcelsius\ to 30\textcelsius\ via the heater requires more than 10 minutes (the latency in the third rule) and the third rule can disable the second rule.
\textit{Group N5} contains 2 SmartApps (4 rules) which involve both rule latency and physical channels (temperature). There is a vulnerability \textbf{V5} that if the temperature is below 18\textcelsius\ when the user presents, the third rule may be activated before the first rule and block its activation.

We summarize the accuracy result of TAPInspector and other approaches in Table \ref{tab:accuarcy}. TAPInspector identifies all known violations in 9 individual SmartApps (\textit{ID 1-9}) and 6 SmartApp groups (\textit{Group 1-6}). Note that \textit{ID 5.1} can generate a fake event using a string-based method invocation by reflection. We fine-tune TAPInspector to support such method invocation to improve the accuracy of ICFG construction. Thus, TAPInspector can successfully identify this fake event leading to turning off the alarm when there is smoke. Besides, TAPInspector also identify violations in our new defined apps (\textit{N1-3} and \textit{Group N4-5}). For these apps, SOTERIA, IoTSAN, and \textsc{IotCom} identified no or incorrect violations. For instance, \textit{N2} unlocks or locks the door depending on the user's presence. IoTSAN and \textsc{IotCom} expose that there is an action conflict in \textit{N2}. But in fact, with different T2A latency, the two actions are scheduled to be executed in order. By considering varying rule latency, TAPInspector identifies the violation in \textit{N2} that the door is unlocked after the user is not present. \textit{Group N4} has a connection-related violation \revise{that other tools cannot find}. TAPInspector considers both the temperature channel as a tardy channel and T2A latency and hence can identify the violation in \textit{Group N5}. But in the detection process of other tools, the first rule can always be executed before the third rule. In summary, TAPInspector achieves 100\% accuracy for this benchmark.

\vspace{-2mm}
\subsection{Vulnerability Detection in Market apps}
\label{sec:detection}
We further evaluate TAPInspector's ability to detect interaction vulnerabilities in market IoT apps. By reviewing safety risks found in existing work \cite{nguyen2018iotsan,ding2018on,birnbach2019peeves}, we first build a set of \textit{safety-sensitive device groups}, including (1) thermostat, air conditioner, heater, and temperature sensor; (2) lock, door control, contact sensor, presence sensor, and motion sensor; (3) smoke detector, carbon dioxide sensor, valve, sprinkler, and water sensor; (4) location mode; (5) security system and alarm; (6) oven and electric blanket. \revise{In each group, there are certain correlations about some attributes between devices, or these devices usually work together to implement specific application scenarios.} We then randomly select apps (2-6 SmartApps and 4-10 IFTTT applets) from 1108 market apps to form an app group. We choose these app groups \revise{that} contain rules associated with a category of device group at least. Besides, we also randomly choose some app groups \revise{containing} switch-capable devices (e.g., switch, outlet) and set a device connection with safety-sensitive devices to simulate misconfigured device connections. \revise{These grouping methods ensure both randomness and practicality since apps installed in a smart home usually have correlations to some extent for device or scene linkage control.}
To configure devices in each app group, we assume a smart home environment containing only one for most devices, but two temperature sensors ($\texttt{Temp}_{in}$ and $\texttt{Temp}_{out}$), two illuminance sensors ($\texttt{Illum}_{in}$ and $\texttt{Illum}_{out}$), and two lights (one in living room and another in bedroom). For these subject names without explicit meanings (e.g., ``myswitch1''), we map them to these devices with the corresponding capability (e.g., light or outlet). Finally, we obtain 204 app groups, and the formal model of each group contains 31 rules \revise{on} average.

To configure preferences (including user preferences and latency in extended actions and tardy attributes) used in apps, we come up with a random but reasonable range for each constant based on common sense regarding how we use the apps. T2A latency has two possible sources: user preferences and platform delay. We configure preference-based latency with a random value up to 20 minutes or the fixed value defined in apps. We do not directly insert platform delays in all rules and just randomly select a small part of them to insert latency. Based on this evaluation environment, TAPInspector totally identifies 533 violations from 204 app groups \revise{by its rule semantic analysis and model checking with properties in Table \ref{tab:property}}. We class them into six major categorizations (F1 to \textbf{F6}), and summarize our findings in Table \ref{tab:vulnerability}.

\begin{table}[!t]
  \caption{Vulnerability detection results with market apps.}
  \label{tab:vulnerability}
  \centering
  \begin{tabular}{l|c|p{4.5cm}}
    \hline
    Categorization & \#violations & Findings \\\hline
    \multirow{3}{*}{\textit{F1}: known}   & \multirow{3}{*}{188} & Action duplication and conflicts are the most two common violations and their root cause is independent application development.\\\hline
    \multirow{4}{*}{\textbf{\textit{F2}}: \textbf{V1}}  &\multirow{4}{*}{67} & Distinguishing immediate and tardy attributes in security analysis of TAP rules is essential. Misconfigured user preference is the major reason of \textbf{V1}. \\\hline
    \multirow{3}{*}{\textbf{\textit{F3}}: \textbf{V2-3}} & \multirow{3}{*}{89}& \textbf{V3} (action overriding) has a different and broader effect than \textbf{V2} (disordered action scheduling).\\\hline
    \multirow{2}{*}{\textbf{\textit{F4}}: \textbf{V4}} & \multirow{2}{*}{41} & Both non-extended and extended actions can break the extended action. \\\hline
    \multirow{4}{*}{\textbf{\textit{F5}}: \textbf{V5-8}} & \multirow{4}{*}{94} & Due to varying platform delays and latency misconfigurations, the number of violations (\textbf{V7-8}) is greater than the one of \textbf{V5-6}.\\\hline
    \multirow{4}{*}{\textbf{\textit{F6}}: \textbf{V9}} & \multirow{4}{*}{61} & No matter how the IoT platform handles offline devices (setting (1) and (2) in $\S$\ref{sec:vuluerability}), there will be such vulnerabilities \textbf{V9}.\\\hline
  \end{tabular}
\end{table}

\textit{F1}: \textit{Action Duplication \& Action Conflicts}. We set violations of these two threats into a single category since there \revise{is} no new vulnerability pattern in them.
TAPInspector pinpoints 136 violations related to action duplication and 52 related to action conflict. We find that rules have similar semantics since in these two threats, thereby TAPInspector can identify these violations only with rule semantic analysis. The main root cause of this category of violation is independent application development.

\textbf{\textit{F2}}: \textit{Action-Trigger Interaction} (\textbf{V1}). From the results of this threat, we find that there are spurious violations identified by rule semantic analysis.
This is because analysis on rule semantics cannot be aware of the behaviors of tardy channel attributes. Hence, by distinguishing immediate and tardy channel attributes in the rule model and setting a random range for each user preference, TAPInspector totally figures out 67 feasible violations with a concrete value space of insecure preferences via model checking.

\textbf{\textit{F3}}: \textit{Action Overriding} (\textbf{V2-3}). With 17 safety and 13 liveness properties (12 are generated according to the properties designed for \textbf{T4} in Table \ref{tab:property}), TAPInspector identifies 81 valid violations associated with action overriding, in which 52 violations come from action conflicts but inserted with random platform delay. We find that between the two types of action overriding, \textbf{V3} has a different and broader effect than \textbf{V2} (disordered action scheduling). The most common violations of \textbf{V3} have two rules having logic similar to the two rules in Fig. \ref{fig:channel} (a), which result in \revise{that} safety-sensitive devices \revise{are} turned on at an unexpected time (e.g., the door is unlocked after the user leaves), or \revise{are} not able to be activated at a specific time (e.g., no alarm when fire).

\textbf{\textit{F4}}: \textit{Action Breaking} (\textbf{V4}). To detect this threat, TAPInspector figures out all TAP rules using extended actions from 1108 apps and sub-models containing the rules whose action may break an extended action. By combining them in a group for each extended action, TAPInspector performs model checking with 18 liveness properties (10 generated via \textbf{T5}) and finally identifies 45 action breaking violations after removing repeated actions. We find that while non-extended actions can break the extended action, extended action is also \revise{possibly} broken by itself. An example violation occurs in two IFTTT applets $r_i$ (``Turn on fan for 15 minutes when $\text{CO}_2>$1000ppm'') and $r_j$ (``At noon turn your fan on for 15 minutes''). If $r_j$ is activated a few minutes (not more than 15 minutes) before $r_i$, the fan will be turned off by $r_j$ first before the extended action of $r_i$ ends, \revise{leading} to security risks.

\textbf{\textit{F5}}: \textit{Condition Blocking} (\textbf{V5-8}). It is also a common vulnerability in our found violations. TAPInspector totally identifies 89 condition blocking violations in the form of \textbf{V5}-\textbf{V8}. In these violations, there are 58 violations related to scheduled condition blocking (\textbf{V7} and \textbf{V8}) which are more than than ones (31 violations) related to condition dynamic blocking (\textbf{V5} and \textbf{V6}). This is caused by varying platform delays and high possible latency misconfigurations. Besides, by setting the four forms of vulnerability in a sequential model (\textit{i.e.}, $r_i$ and $r_j$ are performed in order) and checking manually, we find that they all cannot be identified as actual.

\textbf{\textit{F6}}: \textit{Device Disabling} (\textbf{V9}). In our evaluation, we mainly focus on incorrect connections on outlets. With setting (1) of how an IoT platform handles offline devices (discussed in $\S$\ref{sec:vuluerability}), TAPInspector identifies 32 violations with different devices, including safety-sensitive \textit{actuators} (e.g., siren, door locker, sprinkler) on which related commands cannot be executed, and \textit{sensors} (e.g., motion detector, humidity sensor) on which sensing data is unavailable. With setting (2), TAPInspector identifies 29 violations, in which rules depending on the sensing data of offline devices are not activated as actual. In summary, device disabling is a serious vulnerability \revise{that} is easy to occur since physical and cyber connections depend \revise{highly} on users' manual configurations and \revise{have} a wide range of safety hazards for TAP-based IoT systems.

To evaluate TAPInspector's accuracy for verifying market apps, we randomly select 30 app groups (totally containing 892 rules) to manually examine violations (totally 79 vulnerable rules) found in these groups. Table \ref{tab:at} summarizes the results.
Low precision (88.6\%) and recall (86.4\%) \revise{are} mainly caused by device configurations for rules obtained from IFTTT applets. In the market app set, there are many applets for controlling switch-capable devices, e.g., WeMO switch, SmartThings outlet, D-link SmartPlug, and Mi Home adapter. But we have no configurations about what devices they actually control. Hence, in an app group, we consider them as a single device and may set up a physical connection with a random device (e.g., a heater or a temperature sensor) to simulate misconfigured connections. However, we find that such \revise{a} setting can bring false detection results. On the one hand, these switch-capable devices may be used to control different devices rather than a single device, \revise{resulting} in false positives (3 vulnerable rules) and negatives (4).
On the other hand, these connected devices may also be controllable, \revise{introducing} false positives (6). For example, via the connection with a switch, a heater can be controlled directly by rules, and also indirectly by these rules used to control the switch. Another reason \revise{for} introducing false results is that we do not consider \revise{the} correlation of physical attributes \cite{birnbach2019peeves} when setting channel-based interactions, as shown in Table \ref{tab:interconf}. For example, in fact, turning an air-conditioner on can \revise{affect} both to temperature and humidity, rather than only temperature which results in false negatives (7). A feasible solution to improve TAPInspector's accuracy is to involve more users and vendors to provide practical configurations.

\begin{table}[!t]
  \caption{Accuracy of TAPInspector in our manual examination.}
  \label{tab:at}
  \centering
  \begin{tabular}{c|c|c|c|c}
    \hline
    Random sample size & Precision & Recall & Accuracy  &F-measure \\\hline
    30 (15\%) & 88.6\% & 86.4\% & 98.5\% & 87.5\%\\\hline
  \end{tabular}
\end{table}

\subsection{Performance}

\begin{figure}[!t]
\centering
\includegraphics[width=0.8\columnwidth]{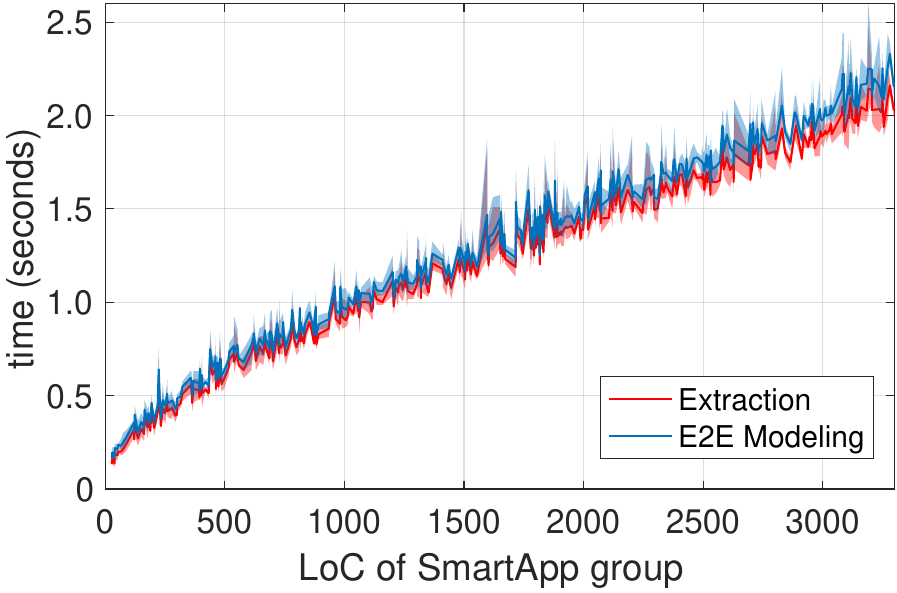}
\caption{Performance of \revise{constructing models for SmartApp groups}.}
\label{fig:extraction}
\end{figure}

\begin{figure}[!t]
\centering
\includegraphics[width=1\columnwidth]{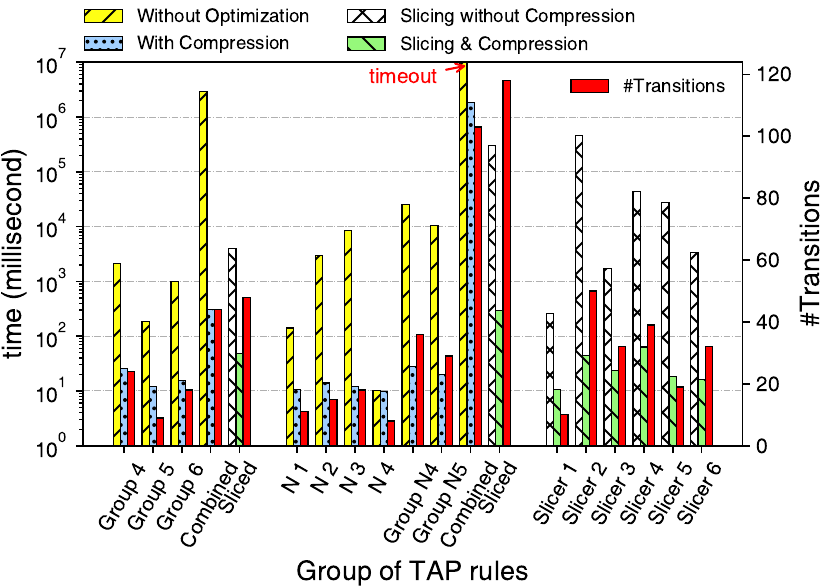}
\caption{Verification time. The right vertical axis represents the number of transitions (\#Transitions) in each model shown on red bars. Others represent model verification time following the left vertical axis.}
\label{fig:performance}
\end{figure}

Finally, we evaluate TAPInspector's performance of vulnerability detection. \revise{We first randomly divided SmartApps (31 used in $\S$\ref{sec:accuracy} and 71 used in $\S$\ref{sec:detection}) into groups of different size, and ran TAPInspector to construct models for these SmartApp groups.
Note that we did not include IFTTT applets in these groups since extracting the rule from each IFTTT applet is always kept within 1 millisecond.}
Fig. \ref{fig:extraction} shows the time of extracting TAP rules from SmartApp groups and the end-to-end time of translating apps' source code into NuSMV models (\textit{i.e.,} E2E modeling). The average E2E modeling time is 1.03 milliseconds per line of code.
We then recorded the analysis time for verifying TAP rules extracted from different apps or app groups, shown in Fig. \ref{fig:performance}. These apps include both the benchmark datasets provided by \textsc{IotCom} (\textit{Group 4-6}) and TAPInspector (\textit{N1-4} and \textit{Group N4-5}), and the market dataset. For each benchmark dataset, we compare the verification time of the individual model of each app or group, their combined model, the model with state compression, and the sliced sub-models. All properties for \textit{Group 4-6} are safety. We can find that the efficiency of verifying the model with model slicing and state compression (48.2 milliseconds (ms)) is significantly improved than verifying original individual models (without optimization is totally 3.3 seconds and with compression is totally 53.2 ms) or the combined model (without optimization is 49.3 minutes and with compression is 307.6 ms).

\begin{table*}[t!]
  \centering
  \caption{Comparison between TAPInspector and related works.}
  \label{tab:com}
  \scalebox{0.85}{
  \begin{threeparttable}
    \begin{tabular}{|l|c|c|c|c|c|c|c|}
    \hline
    Related work & Concurrency & Rule latency & Extended action & Tardy attribute & Channel-based interaction & Connection-based interaction &  Liveness property \\ \hline
    SIFT \cite{liang2015sift}  &   \xmark   &   \xmark     &    \xmark         &    \xmark       &       \xmark   &      \xmark      &    \xmark \\ \hline
    AutoTAP \cite{zhang2019autotap}     &   \xmark   &   \xmark \tnote{$\ddag$}     &    \xmark        &    \xmark       &       \xmark    &      \xmark    &    \xmark   \\ \hline
    IoTGuard \cite{celik2019iotguard}    &   \xmark   &   \xmark     &    \xmark        &    \xmark       &       \xmark    &      \xmark      &    \xmark \\ \hline
    SOATERIA   \cite{celik2018soteria} &   \xmark &   \xmark     &    \xmark         &    \xmark       &       \xmark    &      \xmark     &    \xmark \\ \hline
    IoTSan \cite{nguyen2018iotsan}     &   \cmark\tnote{*}   &   \xmark     &    \xmark       &    \xmark       &       \xmark     &      \xmark      &    \xmark \\ \hline
    IoTMon \cite{ding2018on}     &   \xmark   &   \xmark     &    \xmark            &    \xmark       &       \cmark    &      \xmark  &    \xmark \\ \hline
    \textsc{SafeChain} \cite{hsu2019safechain} &   \xmark   &   \xmark     &    \xmark        &    \xmark       &       \cmark    &      \xmark      &    \xmark \\ \hline
    iRuler  \cite{wang2019charting}    &   \cmark\tnote{$\dag$}  &   \xmark\tnote{$\ddag$}     &    \xmark        &    \xmark       &       \cmark   &      \xmark       &    \xmark \\ \hline
    IoTCom \cite{mohannad2020scalable}     &   \xmark   &   \xmark     &    \xmark         &    \xmark       &       \cmark    &      \xmark     &    \xmark \\ \hline
    HomeGuard \cite{chi2020cross}   &   \xmark   &   \cmark\tnote{$\P$}     &    \xmark        &    \xmark       &       \xmark    &      \xmark      &    \xmark \\ \hline
    TAPInspector         & \cmark  &    \cmark     &     \cmark        &    \cmark       &  \cmark       &     \cmark      &    \cmark \\ \hline
    \end{tabular}
    \begin{tablenotes}
      \footnotesize
       \item[*] Compares a concurrent model without rule latency with a sequential model. \qquad\qquad \qquad \qquad\qquad $\P$ Validates the effect of T2A latency defined by \texttt{runIn}.
       \item[$\dag$] Models IoT systems as a concurrent model, but does not consider concurrency-related issues. \qquad \qquad\quad $\ddag$ Models time only for time-related external events.
     \end{tablenotes}
  \end{threeparttable}}
\vspace*{-4mm}
\end{table*}

For \textit{N1-4} and \textit{Group N4-5}, we perform model checking with liveness properties. The verification time for its sliced models with state compression is totally 295.4 ms, for its individual apps and groups without optimization is 28.2 seconds, but for their combined model without optimization times out. Note that since \textit{N 4} only contains a limited state, its verification time is not explicitly improved through state compression. The third part of Fig. \ref{fig:performance} is about the verification time of market apps. We randomly select 6 model slicers (\textit{Slicer 1-6}) generated by TAPInspector during analyzing market apps and show their verification time with state compression (totally 177.1 ms) or not (totally 8.9 minutes) in Fig. \ref{fig:performance}. We can find that with model slicing and state compression, TAPInspector's verification time is significantly reduced.

\subsection{Discussion and Limitations}

While TAPInspector has been corroborated to be efficient and fast to identify vulnerabilities in TAP rules, it has the following limitations. \textit{First}, more physics should be considered in security analysis, like correlations of physical attributes \cite{birnbach2019peeves}, spatial contexts of different devices, user activity contexts, and sensor-device-user interactions \cite{sikder2021aegis}.
\textit{Second}, \revise{TAPInspector is desired to support runtime detection} since user configurations and physics typically change at runtime. \revise{We can extend our crawler to monitor both IoT app installations and user configuration updates in the IoT system in real time for runtime detection.}
\textit{Third}, our state compression only works on integer attributes. Other types of attributes (e.g., enumerated attributes) should be optimized. \textit{Fourth}, we perform model checking with a random preference space, which is not rigorous. Implementing an entire preference space analysis should further improve TAPInspector' analysis ability.
\textit{Fifth}, our security analysis only focused on the impact of rule semantic and physical attributes. The next step is to involve more features of IoT architecture, e.g., communication and control layers, to enable a cross-layer and systematic analysis.

\revise{Furthermore, we find that the major root causes of vulnerabilities in TAP rules come from various aspects, including user configurations, IoT platforms, and logic in IoT apps. It is hard to address these vulnerabilities by a vendor individually. The current consensus in academia it to establish a mechanism that can automatically detect and repair them. Such a mechanism can help users fix TAP rules (e.g., \cite{zhang2019autotap}) or enforce TAP rules following security and safety policies (e.g., \cite{celik2019iotguard}). We believe that extending TAPInspector as an end-to-end tool for detecting and repairing vulnerabilities is more valuable.}

\section{Related Work}

IoT security has received wide \revise{attentions} from several aspects \cite{alrawi2019sok}, such as IoT devices \cite{kumar2019all}, protocols \cite{ronen2017iot}, side channels \cite{han2018you}, and platforms \cite{manandhar2020towards}. Here, we mainly discuss the efforts related to our research. To ensure the security of IoT apps, analyzing their source codes and detecting violations with static analysis is a much more powerful approach, which has been widely followed, including SOATERIA~\cite{celik2018soteria}, IoTSAN~\cite{nguyen2018iotsan}, HomeGuard~\cite{chi2020cross}, \textsc{IotCom}~\cite{mohannad2020scalable}, and also our TAPInspector. SOATERIA \cite{celik2018soteria}, HomeGuard~\cite{chi2020cross} and \textsc{IotCom}~\cite{mohannad2020scalable} are similar in extracting behavior models from IoT apps that they perform AST analysis on the Groovy codes of APPs and construct ICFGs for model construction. Since apps in many IoT platforms \cite{ifttt, smartthings, openhab} are open-source, \revise{this} approach can provide finer-grained and deterministic meanings for model construction. We also follow this advanced approach. IoTSAN~\cite{nguyen2018iotsan} translates the Groovy code of apps to Java for using the Bandera \cite{hatcliff2001using}, a toolset for model checking Java programs. This translation eases the model construction but can also result in limited semantics for covering new apps.

Different from source code analysis of IoT apps, IoTMon~\cite{ding2018on} and iRuler~\cite{wang2019charting} present \revise{an} approach that leverages NLP techniques to identify behavior models from IoT apps for model checking or risk analysis. Such an NLP-based approach provides the support security analysis for where source codes are not available. \textsc{SafeChain}~\cite{hsu2019safechain} is a model-checking-based system for privilege escalation and privacy leakage. It constructs an FSM of IoT systems manually and automatically optimizes the FSM's state space by grouping functional equivalent attributes and pruning redundant attributes. Such a model optimization can also work on our models to further reduce verification time.

SIFT~\cite{liang2015sift}, AutoTAP~\cite{zhang2019autotap}, ContexIoT~\cite{jia2017contexiot}, and IoTGuard~\cite{celik2019iotguard} focus on action-related violation detection and resolution in a static \cite{liang2015sift,zhang2019autotap} or dynamic \cite{jia2017contexiot, celik2019iotguard} manner. SIFT~\cite{liang2015sift} is a safety-centric programming platform that detects action conflicts and provides assistance in understanding how to correct problems through model checking. AutoTAP~\cite{zhang2019autotap} is a tool that can automatically improve or generate TAP rules to repair flawed rules specified by users according to given properties. For dynamic resolution, ContexIoT~\cite{jia2017contexiot} analyzes individual IoT apps to prevent sensitive information leakage at run-time. IoTGuard~\cite{celik2019iotguard} is a dynamic rule enforcement system that can prevent the executions of these actions leading to violations. Violation detection in these works is limited to action conflicts and duplication. Therefore, automated violation resolution is worthy of further study for more comprehensive violations.

To compare our work against these above techniques, we present their comparison over various features in Table \ref{tab:com}. Rule interaction and physical channel have been widely studied. For other features, in IoTSAN \cite{nguyen2018iotsan}, the authors discussed the difference between concurrency and sequential model of IoT systems. They experimented with small systems and found that the sequential approach can discover all violations found \revise{by} the concurrent one. They did not consider rule latency and assumed that the internal events associated with an external event are handled atomically in order. Therefore, their concurrent model is much similar to the sequential one. iRuler \cite{wang2019charting} models IoT apps in a concurrent model, and also models time as a monotonically increasing variable for time-related events. However, this work did not study concurrency-related issues. To the best of our knowledge, TAPInspector is the first technique for automated security analysis of IoT apps with concurrency, rule latency, extended actions, tardy-channel-based interactions, connection-based interactions, and liveness properties. That is why TAPInspector can identify several new types of interaction vulnerability.

\section{Conclusion}
In this work, we define 9 new types of rule interaction vulnerabilities in TAP-based IoT systems and design TAPInspector, a novel automatic vulnerability detection tool for TAP-based IoT systems. With a comprehensive analysis of the TAP-based IoT system, we identify several critical features which motivate us to design the model-checking-based security analysis system to inspect concurrent IoT systems. With automatic rule extraction and model construction, TAPInspector addresses challenges in inspecting the concurrent IoT system, such as formulizing TAP rules with various rule latency and reducing state space of models due to concurrency and rule latency. By verifying with a set of safety and liveness properties, our evaluations show that TAPInspector can identify violations from market IoT apps with an excellent performance overhead.

\bibliographystyle{IEEEtran}
\bibliography{IEEEabrv,reference}

\end{document}